\documentclass[epj]{svjour} 
\usepackage{epsf,cite,epsfig,psfig,float,amssymb,stmaryrd,latexsym} 
\usepackage{graphics,psfrag}
\usepackage{graphicx} 
\newcommand{\be}{\begin{equation}} 
\newcommand{\ee}{\end{equation}} 
\newcommand{\beqa}{\begin{eqnarray}} 
\newcommand{\eeqa}{\end{eqnarray}}

\begin{document} 
\titlerunning{Infrared regularization for spin-1 fields}
\title{Infrared regularization for spin-1 fields
\thanks{Work supported in part by   
Deutsche Forschungsgemeinschaft through grants provided for the SFB/TR 16
``Subnuclear Structure of Matter''.}} 
 
\author{Peter C. Bruns \inst{1} \thanks{Electronic address:~bruns@itkp.uni-bonn.de}  
\and 
Ulf-G. Mei{\ss}ner \inst{1}\inst{2} \thanks{ 
Electronic address:~meissner@itkp.uni-bonn.de}  
}                     % Do not remove 
% 
%\offprints{Ulf-G. Mei{\ss}ner}          % Insert a name or remove this line 
% 
\institute{  
Helmholtz-Institut f\"ur Strahlen- und Kernphysik (Theorie), 
Universit\"at Bonn, Nu{\ss}allee 14-16, D-53115 Bonn, Germany 
\and 
Forschungszentrum J{\" u}lich, Institut f{\" u}r Kernphysik 
(Theorie),  D-52425  J{\" u}lich, Germany 
} 
\date{Received: date / Revised version: date} 
\abstract{ 
We extend the method of infrared regularization to spin-1 fields. As applications,
we discuss the chiral extrapolation of the rho meson mass from lattice QCD data 
and the pion-rho sigma term.
 % end of PACS codes 
} %end of abstract
\maketitle 

\tableofcontents

%%%%%%%%%%%%%%%%%%%%%%%%%%%%%%%%%%%%%%%%%%%%%%%%%%%%%%%%%%%%%%%%%%%%%%%%%%%%%%
\section{Introduction} 
\def\theequation{\arabic{section}.\arabic{equation}} 
\label{intro}

Chiral perturbation theory (CHPT) is the effective field theory of the
Standard Model at low energy \cite{GL1,GL2}. It is based on the spontaneously broken
approximate chiral symmetry of QCD. The pions, kaons and the eta can be
identified with the Goldstone bosons of chiral symmetry breaking. Their
interactions are weak and vanish in the chiral limit of zero quark masses
when the energy goes to zero. This is a consequence of Goldstone's theorem
and allows for a consistent power counting. Consequently, any amplitude can be written
as sum of terms with increasing powers in external momenta and quark masses,
symbolically
\begin{equation} 
{\mathcal A} = p^D \, f\left( p/\mu , g\right)~,
\end{equation}
where $f$ is a function of order one,
$p$ collects the small parameters, $D$ is the chiral dimension, $\mu$ a 
regularization scale (related to the UV divergences in loop graphs)
and $g$ a collection of coupling constants, the so-called
low-energy constants (LECs).  Weinberg \cite{Wein}  first established this power
counting, in particular, the expansion in $p$ (the chiral expansion)
can be mapped onto an expansion in terms
of tree and loop graphs, with $n$ loop graphs being parametrically suppressed by
powers of $p^{2n}$ compared to the leading trees. The explicit expression for $D$
reads:
\begin{equation}\label{chiraldim}
D({\mathcal A}) = \sum_N V_n (n-2) + 2L +2~,
\end{equation}
with $L$ the number of Goldstone boson loops and $V_n$ a vertex of 
order ${O}(p^n)$.
In essence, this power counting
works because the pion mass vanishes in the chiral limit and thus the only dimensionfull
scale in this limit is the pion decay constant $F_\pi$ 
(more precisely, its value in the chiral limit). 
Utilizing a symmetry preserving regularization scheme
like e.g. dimensional regularization leads to homogeneous functions in the small
parameters (for more details, see the reviews \cite{ulf,pich,ecker,BKMrev,scherer}). 
The precise relation between  this effective field theory (EFT) and the chiral Ward 
identities of QCD was firmly 
established in  Refs.~\cite{Heiri,SWMIT}. 

The active degrees of freedom in CHPT are the Goldstone bosons, chirally coupled to
external sources. It is, however, known since long that vector and axial-vector 
mesons also play an important role in low energy hadron physics, as one example
we mention the fairly successful description of the pion charge form factor in
terms of the (extended) vector dominance approach (for more details, see e.g. the
reviews \cite{UlfV,KoichiV}). These heavy degrees of freedom decouple in the chiral
limit and at low energy from the Goldstone bosons. Nevertheless, they leave their
imprint in the low-energy EFT of QCD by saturating the LECs, which has been termed
resonance saturation \cite{NuB,Donoghue,Eck}. We will discuss this issue briefly 
in Sect.~\ref{sec:tree}. Here, we are interested in an extension of CHPT where these
spin-1 fields are accounted for explicitly. While it is straightforward to construct
the corresponding chiral effective Lagrangian (as briefly reviewed also
in Sect.~\ref{sec:tree}), the computation of loop diagrams
is not. This is related to the appearance of a large mass scale in this EFT,
namely the non-vanishing chiral limit mass of the spin-1 fields. This scale destroys the
one-to-one mapping between the chiral expansion and the loop expansion, as discussed
in more detail in Sect.~\ref{sec:problem}. To be able to proceed in a systematic fashion,
one has to be able to separate the contributions to loop diagrams originating from this 
large mass scale in a controllable and symmetry-preserving fashion. This problem also
appears in the EFT when nucleons (baryons) are included (in fact, it has been analyzed
first in this context \cite{GSS}), and various solutions have been
suggested, like heavy baryon CHPT \cite{JM,BKKM}, subtraction schemes for the hard
momenta \cite{Tang}, infrared regularization \cite{BL} or extended on-mass shell
regularization \cite{Mainz1}. For the case of the vector mesons considered here
an additional complication arises - these particles can decay into Goldstone bosons
and thus appear in loops without appearing in external lines. Here, we will
present an extension of infrared regularization that allows to treat such diagrams. 
In Sect.~\ref{sec:softhard}, we briefly review the intuitive approach of Ref.\cite{Tang}
how the contributions from the hard scale can be tamed. The more elegant infrared
regularization \cite{BL} is introduced in Sect.~\ref{sec:IR}. In Sect.~\ref{sec:IRnew},
we discuss the new contributions to the Goldstone boson
self-energy graph when the heavy particle
only appears in the loop. The singularity structure of these loop graphs is analyzed in
Sect.~\ref{sec:sing}. Based on that, we show how the infrared singular part can
be obtained for such type of one-loop diagrams in Sect.~\ref{sec:IRcorr}. The method is then
applied to the self-energy graph where the spin-1 field only appears inside the
loop, see Sect.~\ref{sec:self1}. The corresponding
triangle graph is discussed in Sect.~\ref{sec:tria}.
Section~\ref{sec:Vself} contains the analysis of the vector meson self-energy 
diagram with a pure Goldstone boson loop, which does not only involve soft momenta.
As an application, we discuss the chiral extrapolation of lattice QCD data for
the rho meson mass and related topics
in Sect.~\ref{sec:mrho}. A brief summary is given in  Sect.~\ref{sec:summ}.
Some technicalities are relegated to the appendices. For other works on the problem
of vector mesons in chiral EFT, we refer 
to \cite{JM95,BM95,Urech,bijnens,PichV,Koichi,MainzV}.

%%%%%%%%%%%%%%%%%%%%%%%%%%%%%%%%%%%%%%%%%%%%%%%%%%%%%%%%%%%%%%%%%%%%%%%%%%%%%%%%%%%%%
\section{Prelude: Vector mesons in trees}
\setcounter{equation}{0} 
\label{sec:tree}
In this section we show how vector mesons can be treated in chiral
perturbation theory when only tree graphs are considered. This material is not
new, but is needed to set up the formalism and to set the stage for the
discussion of vector mesons in loops. The reader familiar with this material
might skip this section. Also, our considerations are more general, they
really refer to the coupling of heavy degrees of freedom to the Goldstone
boson fields. Furthermore, when talking about vector mesons, we really mean vector
{\em and} axial-vector mesons (spin-1 fields).

Our aim is to write down an effective Lagrangian containing the vector meson
resonances explicitly. The word 'explicitly' refers to the fact that the
vector meson resonances are present {\em implicitly} in the Goldstone boson
effective Lagrangians through their contributions to (some of) the pertinent 
low-energy constants, see Refs.~\cite{NuB,Donoghue,Eck}.
We want to state this on a more formal level. Assume that we have constructed a Lagrangian ${\mathcal L}_{Res}(R,U,v,a,s,p)$ where $R$ are some resonance fields which might be the vector mesons for example. $U$ collects the Goldstone bosons and $v,a,s,p$ are vector,
axial-vector, pseudoscalar and scalar sources. The latter also include the quark
masses, $s(x) = {\mathcal M}$, with ${\mathcal M} = {\rm diag}(m_u,m_d,m_s)$ the
quark mass matrix. The resonances are all very much heavier than the Goldstone bosons (e.g. $M_{\rho} \sim 770$ MeV), and therefore it is a consistent procedure for a low-energy effective theory to integrate out these heavy degrees of freedom  by means of a path integration over $R$:
\begin{equation}
\int [dR] e^{i\int d^{4}x {\mathcal L}_{Res}(R,U,v,a,s,p)} = e^{iZ_{ind}(U,v,a,s,p)} .
\end{equation}
By doing the path integral, a Goldstone boson theory (containing only $U$ and
the external fields) is recovered. $Z_{ind}(U,v,a,s,p)$ may be called the
generating functional {\em induced} by 
${\mathcal L}_{Res}(R,U,v,a,s,p)$ through the path integration. Such a step may be visualized in a Feynman graph by shrinking the lines symbolizing the $R$-propagators to a point, and attaching contact terms (interactions between the remaining fields) to these points.

What is the physical content of all this? Computing the $R$-induced Goldstone boson field theory means that some interaction terms between the Goldstone (and external) fields are computed by the path integration over $R$ and are therefore expressed through couplings of the resonance to these fields and the resonance mass $M_{R}$, which can be measured in processes where the resonance occurs as an external state. The interactions generated in this way can be compared with the couplings determined by the LECs of the original Goldstone boson field theory. This will show us how important the resonance contributions to the processes under consideration are. We get a microscopic information on these processes which we could not achieve with the theory of the light fields alone.

Numerically, however, one could as well work with the original effective field theory and simply include higher and higher orders. As should be clear by now, the resonance contributions are {\em implicitly} present in this theory, influencing the values of the LECs.
This can be illustrated by considering a diagram with a resonance line through which a small ($O(p)$) momentum flows. The resonance propagator can be expanded in this case using
\begin{equation}
\frac{1}{p^{2} - M_{R}^{2}} = -\frac{1}{M_{R}^{2}}\biggl( 1+\frac{p^{2}}{M_{R}^{2}}+ \ldots \biggr)~,
\end{equation}
leading to an infinite series generating terms of an arbitrary high order. Including the resonance field explicitly, one takes care of {\em all} terms in this series and not just the first few terms of it. This can be advantageous. A nice example can be found in \cite{Kubis}, where the inclusion of vector mesons substantially improves the results for the nucleon form factors computed in that work.

With this motivation, let us now concentrate on the vector mesons. Following
\cite{NuB}, we will (first) use an antisymmetric Lorentz tensor-field $W_{\mu
  \nu} = - W_{\nu \mu}$ to describe the vector meson. This has six degrees of
freedom, but we can dispose of three of them in a systematic way, for details
see App.~\ref{app:ten} and Ref.~\cite{NuB}. 
The spin-1 fields transform as any matter field under the
non-linearly realized chiral symmetry,
\begin{equation}
W_{\mu\nu}(x) \rightarrow h W_{\mu\nu}(x) h^{\dagger}~ ,
\end{equation} 
where the compensator field $h$ is defined via
\begin{equation}
u(x)\rightarrow g_{R} u(x) h^{\dagger} = h u(x) g_{L}^{\dagger}
\end{equation}
with $g_I$ is an element of SU(3)$_I$, $I=L,R$ and $u^2 = U$.
The kinetic and mass term of the effective
Lagrangian for vector mesons has the form
\begin{equation}\label{Lkin}
{\mathcal L}_{W}^{kin} = 
-\frac{1}{2}\langle \nabla^{\mu}W_{\mu \nu}\nabla_{\rho}W^{\rho \nu}\rangle 
+ \frac{1}{4}M_{V}^{2}\langle W_{\mu \nu}W^{\mu \nu}\rangle~,
\end{equation}
where 
\begin{equation}\label{WT}
W_{\mu \nu} = \frac{1}{\sqrt{2}}W_{\mu \nu}^{a}T^{a}
\end{equation}
for an octet of vector mesons, and summation over the flavor index $a=1,
\ldots , 8$ is implied, and $\langle \,\, \rangle$ denotes the trace in flavor space.
The pertinent covariant derivative is
\begin{equation}
\nabla^{\mu}R = \partial^{\mu}R + [\Gamma^{\mu}, R]~. 
\end{equation}
It transforms as $W_{\mu\nu}$ under the chiral group. Here, $\Gamma^{\mu}$ is the connection,
\begin{equation}
\Gamma^{\mu} = \frac{1}{2}(u^{\dagger}[\partial^{\mu}-i(v^{\mu}+a^{\mu})]u 
+ u[\partial^{\mu}-i(v^{\mu}-a^{\mu})]u^{\dagger})\, .
\end{equation}
For the $SU(3)$ case we consider here, the $T^{a}$ are the usual
Gell-Mann-matrices which obey $\langle T^{a}T^{b}\rangle = 2\delta^{ab}$ as well as
$[T^{a},T^{b}] = 2if^{abc}T^{c}$,
where $f^{abc}$ are the totally antisymmetric structure constants of $SU(3)$.
The mass $M_V$ appearing in Eq.(\ref{Lkin}) is strictly speaking the vector meson
mass in the chiral limit.

The vector meson octet we consider here may be given in matrix form:
\begin{equation}
\mathbf{W} = \left( \begin{array}{ccc}
\frac{\rho^{0}}{\sqrt{2}}+\frac{\omega_{8}}{\sqrt{6}} & \rho^{+} & K^{\ast+} \\
\rho^{-} & -\frac{\rho^{0}}{\sqrt{2}}+\frac{\omega_{8}}{\sqrt{6}} & K^{\ast0} \\
K^{\ast-} & \bar K^{\ast0} & -\frac{2\omega_{8}}{\sqrt{6}}
\end{array} \right)~.
\end{equation}
From the above Lagrangian, one can derive the propagator, 
\begin{eqnarray}
\langle 0\mid T(W_{\mu \nu}^{a}(x)W_{\rho \sigma}^{b}(y))\mid 0\rangle = 
\frac{i\delta^{ab}}{M_{V}^{2}}\int
\frac{d^{4}k}{(2\pi)^{4}}\frac{e^{ik(x-y)}}{M_{V}^{2}-k^{2}-i\epsilon} \nonumber\\
\times [g_{\mu \rho}g_{\nu \sigma}(M_{V}^{2}-k^{2})+g_{\mu
  \rho}k_{\nu}k_{\sigma}-g_{\mu \sigma}k_{\nu}k_{\rho} - (\mu \leftrightarrow
\nu)]\, . &&\nonumber \\ &&
\end{eqnarray}
Now we examine the interaction of the vector meson field with the other fields of the theory, especially the Goldstone boson fields. We actually already have interaction terms coming from the covariant derivative, but they are bilinear in $W_{\mu \nu}$. For the simplest diagrams we want to consider, we need vertices with only one vector meson line attached to them, i.e. couplings linear in $W_{\mu \nu}$. Therefore we neglect the 'connection terms' in the following.
From the philosophy of effective field theories, we are required to construct the most general interaction terms consistent with Lorentz invariance, chiral symmetry, parity, charge conjugation and hermiticity. The building blocks with which this can be done are, in principle, at hand: the Goldstone boson fields (collected in $U$ or its square root $u$), the covariant derivative $D^{\mu}$ acting on $U$, the object $\chi$ which contains the scalar and pseudoscalar fields (in particular, the quark mass matrix), the field $W_{\mu \nu}$ (and the pertinent covariant derivative $\nabla_{\mu}$ acting on it) and the field strength tensor $F_{\mu \nu}$ for the external fields $v_{\mu}$ and $a_{\mu}$.
For our purposes, it is more convenient  to collect these blocks in the combinations 
\begin{eqnarray*}
u_{\mu} &=& iu^{\dagger}D_{\mu}Uu^{\dagger} = u_{\mu}^{\dagger}~, \\
u_{\mu \nu} &=& iu^{\dagger}D_{\mu}D_{\nu}Uu^{\dagger}~, \\  
\chi_{\pm} &=& u^{\dagger}\chi u^{\dagger} \pm u\chi^{\dagger}u~, \\
F^{\pm}_{\mu \nu}&=&uF^{L}_{\mu \nu}u^{\dagger}\pm u^{\dagger}F^{R}_{\mu \nu}u~.
\end{eqnarray*} 
This is better from a practical point of view because the so-defined objects all transform like $W_{\mu \nu}$ under chiral transformations. This makes it easy to find chirally invariant expressions: Just take some of the above objects and put them inside a trace $\langle \ldots \rangle$. This will then be invariant by the cyclicity property of the trace. Of course, one can further reduce these possibilities by imposing the other symmetries mentioned above, and using the antisymmetry of $W_{\mu \nu}$.

What concerns power counting, it is clear that $u^{\mu}$ is a quantity of order $O(p)$ due to the covariant derivative $D^{\mu}$ giving one factor of momentum or external (axial-)vector source. Similarly, the other objects in the list are of $O(p^{2})$.

At order $O(1)$, there is no term linear in $W$ because $W_{\alpha \alpha}=0$.
The lowest order interaction terms turn out to be of order $O(p^{2})$ and read \cite{NuB}
\begin{equation}\label{Wint}
L^{int}_{W} = \frac{F_{V}}{2\sqrt{2}}\langle F^{+}_{\mu \nu}W^{\mu \nu}\rangle + \frac{iG_{V}}{2\sqrt{2}}\langle [u_{\mu},u_{\nu}]W^{\mu \nu}\rangle .
\end{equation}
This is more complicated than it looks, because both terms contain interactions with an arbitrary high (even) number of Goldstone boson fields. One must carefully expand the objects $F^{\pm}_{\mu \nu}$ and $u_{\mu}$ to obtain the vertex for a particular amplitude.
It is now clear why vector meson singlets can be neglected: The sources to
which they would be allowed to couple are $\langle F^{\pm}_{\mu \nu}\rangle$
and $\langle [u_{\mu},u_{\nu}]\rangle$, but both traces are zero.
 
%One remark concerning the numerical values of $F_V$ and $G_V$ : We do not consider the 
%vector mesons as gauge bosons of chiral symmetry. This ansatz would lead to 
%the relation $F_{V} = 2G_{V}$ \cite{Eck}. This will not be assumed here, though it is 
%fulfilled quite well in nature: From the measured decay rates for $\rho^{0}\rightarrow e^{+}e^{-}$ 
%and $\rho \rightarrow 2\pi$, one has $F_{V} = 154 ~MeV$ and $G_{V} = 69 ~MeV$. 
%The last value will have to be corrected slightly because the pions from the 
%rho decay are not soft. In \cite{NuB}, the value of $G_{V}$ is taken from the 
%assumption of vector meson dominance concerning the form factor of the Goldstone 
%boson. This will mean that the the number $F_{V}G_{V}$ will be taken as an 
%experimental input, which together with the measured value of $F_{V}$ 
%determines $G_{V} = 53 ~MeV$. 
Note that a discussion of the numerical values of $F_V$ and $G_V$ is given 
in Refs.~\cite{NuB,Eck}. Furthermore, the  vector field formulation is 
summarized in App.~\ref{app:vec}.

%%%%%%%%%%%%%%%%%%%%%%%%%%%%%%%%%%%%%%%%%%%%%%%%%%%%%%%%%%%%%%%%%%%%%%%%%%%%%%%%%%%%%%%%%%%%%%
\section{Vector mesons in loops: Statement of the problem}
\setcounter{equation}{0} 
\label{sec:problem}
Difficulties arise when in a Feynman diagram lines representing a heavy matter field are part of a loop. The corresponding amplitude will in general not be of the chiral order expected from power counting. This has been noted many years ago when the nucleon was incorporated in CHPT \cite{GSS}.
We already saw in the last chapter that power counting in CHPT is not as straightforward as in the purely Goldstone bosonic sector, and we already noted the reason for this, namely, the presence of a new large scale: The mass of the heavy matter field.
In \cite{GSS}, it was shown that the parameters of the lowest order pion-nucleon Lagrangian already were (infinitely) renormalized by loop graphs in the chiral limit. There exists a mismatch between the loop expansion in $\hbar$ and the chiral expansion in small parameters of order $O(p)$. The loop graphs in general generate power counting violating terms, confusing the perturbative scheme suggested by CHPT. For example, a graph with dozens of loops might give a contribution as low as $O(p^{2})$. Clearly, we will have to get rid of these power counting violating terms if we want to keep this scheme when including heavy matter fields.

For graphs where the heavy field only shows up in internal tree lines, the problem concerning power counting was not urgent, because the chiral expansion of the amplitude corresponding to such a graph at least {\em started} with the correct order. We saw this in the last chapter: We counted the vector meson propagator as $O(1)$, and in the low-energy region, the momentum transfer variable $t$ was much smaller than the square of the heavy mass, allowing an expansion of the propagator in the small dimensionless variable $t/M_{V}^{2}$, which starts at $O(1)$.

%%%%%%%%%%%%%%%%%%%%%%%%%%%%%%%%%%%%%%%%%%%%%%%%%%%%%%%%%%%%%%%%%%%%%%%%%%%%%%%%%%%%%%%%%%%%%%
\section{Soft and hard poles}
\setcounter{equation}{0} 
\label{sec:softhard}
If the heavy particle line belongs to a loop, an integration over the four-momentum flowing through this line takes place. Due to the pole structure of the propagator, the integral will pick up a large contribution from the region where the line momentum squared $k^{2} \approx M_{V}^{2}$, with $M_{V}$ the heavy mass (we keep the index $V$ to remind us that we will concentrate on vector mesons, but the present discussion is more general). This is the region of the 'hard poles' in the terminology of \cite{Tang}. These contributions are of high-energy origin and clearly do not fit in a low-energy effective theory - they must be identified as generating the part of the loop integral which violates the power counting scheme, because this scheme would clearly be valid if these 'hard poles' were missing. The only hard-momentum effects for loop integrals in the Goldstone boson sector are the ultraviolet divergences, which are handled by dimensional regularization.  

Before discussing the infrared regularization method, for illustrative purposes
we will shortly present the idea of \cite{Tang}. We remarked that that the power counting violating terms stem from the region $k^{2}\approx M_{V}^{2}$. Far off that region, for $k^{2} \ll M_{V}^{2}$, it would be allowed to expand the propagator in the small variable $k^{2}/M_{V}^{2}$, as we did for tree lines in the last section. Of course, this is not allowed under a momentum integral which extends to arbitrary high momenta $k^{\mu}$. But it is too much of a temptation to do so, {\em because in this way one destroys the 'hard poles' responsible for the power counting violating terms}. Doing the expansion, treating the loop-momentum $k^{\mu}$ over which one integrates as a small quantity, and interchanging integration and summation of this expansion, one ends up with an expression which obeys power counting, because the hard poles are not present in any individual term of the series which one integrates term by term. Only the soft poles from the Goldstone boson (or perhaps also photon) propagators will be present in the individual integrals.

Clearly, this is not the old integral any more, but a certain part of it, collecting only the contributions from the 'soft poles' - it is called the 'soft part' of the full integral in the terminology of \cite{Tang}. The remaining part, which was dropped by this procedure, collects the contributions from the 'hard poles' and stems from the high-energy-momentum region. It is argued that this part is expandable in the small chiral parameters and, truncated at a sufficiently high order (depending on the order to which one calculates) it is a local polynomial in these small parameters and can be taken care of by a renormalization of the parameters of the most general effective Lagrangian. The power counting violating terms are then hidden in the renormalization of the LECs, and the 'soft part', with subtracted residual high-energy divergences, is taken as the renormalized amplitude appearing as a part of the perturbation series. If this argument is correct, and the part of the full integral which is dropped is indeed expandable in the small parameters, the 'soft part' contains all the terms of the full integral which are non-analytic in the expansion parameters, like terms of the typical form 
$\ln M_{\phi}/\mu$, the so-called 'chiral log'-terms where $M_{\phi}$ is the Goldstone boson mass,
and $\mu$ is the renormalization scale (we will use dimensional regularization throughout).
This becomes large if one lets the Goldstone boson mass go to zero. The physical interpretation is that in this limit the range of the interaction mediated by the Goldstone bosons becomes infinite, so that, for example, the scalar radius of the pion (or other hadrons) is infrared divergent in the chiral limit, which makes sense because it measures the spatial extension of the Goldstone boson cloud. 

We keep in mind that the 'soft poles' in the region where the loop momentum $k\sim O(p)$ are responsible for the terms non-analytic in the small parameters, like chiral log terms, and that all these terms {\em obey a simple power counting}. This follows from the argument given by \cite{Tang}. If one ever encounters a power counting violating term non-analytic in the quark masses, for example, this would invalidate the above argument - such terms can not be hidden in the renormalization of the parameters of an effective Lagrangian.

We would like to demonstrate that arguments such as the one cited above are indeed more than just wishful thinking. The basic features of the argument can be exhibited by a toy model loop integral (it is straightforward to generalize it -  the formulas would only look a bit more complicated).

Consider the integral
\begin{equation}
I = \oint_{C} \frac{dk}{2\pi i} \frac{f(k)}{(k-a)(k-b)}
\end{equation}
where $C$ is some contour in $\mathbf{C} \setminus \{a,b\}$ (enclosing the poles or not) and $f(k)$ is an analytic function. Such an integral  might e.g. be the $k^{0}$-integration over a full loop integral. The soft pole is called $a$ and is of $O(p)$ while the hard pole is $b$. In fact, we will only need $|a| < |b|$.
In the spirit of the argument of \cite{Tang}, we destroy the hard pole in two steps 
\begin{eqnarray}\label{soft}
I &\rightarrow& \oint_{C} \frac{dk}{2\pi i}\frac{f(k)}{(k-a)}
\left(-\frac{1}{b}\right)
\left( 1+\frac{k}{b} + \frac{k^{2}}{b^{2}} + \ldots \right)\nonumber \\
&\rightarrow& -\frac{1}{b}\sum_{n=0}^{\infty}\oint_{C}\frac{dk}{2\pi i}\frac{f(k)}{(k-a)}\frac{k^{n}}{b^{n}}\nonumber \equiv I_{\rm soft}~.
\end{eqnarray}
Depending on the form of $f(k)$, this will be ultraviolet divergent if the contour extends to infinity, but this divergence has nothing directly to do with the pole structure and can be handled by a regularization scheme.
By the method of residues, $I_{\rm soft}$ is computed to be
\begin{equation}
I_{\rm soft} = \frac{f(a)}{(a-b)}\oint_{C}\frac{dk}{2\pi i}\frac{1}{(k-a)} .
\end{equation}
We were allowed to sum the geometric series because $|a| < |b|$. Therefore $I_{\rm soft}$ is just the first summand in the decomposition
\begin{eqnarray}
I &=& \oint_{C}\frac{dk}{2\pi i}\frac{f(k)}{(k-a)(k-b)}  = \nonumber\\ 
&=& \frac{1}{a-b}\biggl(\oint_{C} \frac{dk}{2\pi i}\frac{f(a)}{(k-a)} 
- \oint_{C}\frac{dk}{2\pi i}\frac{f(b)}{(k-b)}\biggr) \nonumber \\ 
&=& I_{\rm soft} + I_{\rm hard}~,
\end{eqnarray}
which clearly separates the contributions from the soft and the hard pole, respectively. We have thus demonstrated that the expansion of the hard pole structures followed by an interchange of summation and integration indeed gives {\em exactly} the contribution from the soft pole. Note that the above argument can be iteratively used for multiple poles.

A complementary approach would be to treat $k$ as large, expand the soft pole structure in the small variable $a$ and again interchange summation and integration, thereby isolating the hard pole contribution which is by construction analytic in the small parameter $a$. This may be called the 'regular' part of the loop integral. Again, divergences can show up which have nothing to do with the details of the pole structure (infrared divergences if the contour $C$ encloses $k=0$), but apart from this, the calculation analogous to the preceding one will give nothing but $I_{\rm hard}$. In this sense, the approaches of \cite{Tang} and \cite{Mainz1}  can be said to be complementary to each other.

We will see in the next section that the arguments of \cite{Tang} can be refined or, as one should say, modified in a rather elegant way. In particular, the method
described in this section does not always work because not all integrals converge
in the low energy region, for more details on that issue, see e.g. \cite{BL}.
%Before closing this section, I want to refer to \cite{HeavyBaryon, BKM92}, where other clever ideas have been developed to incorporate heavy matter fields in $CHPT$ and which were found earlier than the kind of treatment cited above. Much work has been done in that direction, but this is not the place to describe these methods.  

%%%%%%%%%%%%%%%%%%%%%%%%%%%%%%%%%%%%%%%%%%%%%%%%%%%%%%%%%%%%%%%%%%%%%%%%%%%%%%%%%%%%%%%%%%%%%%%%%%%%%
\section{Infrared regularization}
\setcounter{equation}{0} 
\label{sec:IR}
In order to find a more elegant way to separate the low-energy part of the loop integrals, we will examine a loop consisting of one 'heavy' propagator and one Goldstone propagator. We use dimensional regularization to handle the ultraviolet divergence of such an integral, i.e. we give a meaning for the notion of an integral in $d$ dimensions, where $d$ might be fractional, negative , etc.$\,$.

We consider the scalar loop integral
\begin{equation}\label{IVphidef}
I_{V\phi}(q) = i\int \frac{d^{d}k}{(2\pi)^{d}}\frac{1}{((k-q)^{2}-M_{\phi}^{2})(k^{2}-M_{V}^{2})} .
\end{equation}
Here $q$ denotes an external momentum flowing into the loop, $M_{\phi}$ is the Goldstone boson mass and $M_{V}$ is the mass of the heavy particle. The '$i\epsilon$'-prescription, giving the masses a small negative imaginary part, should be understood here. We leave it out because it will play no role in the following discussion.

\begin{figure}[htb]
\centerline{\psfig{file=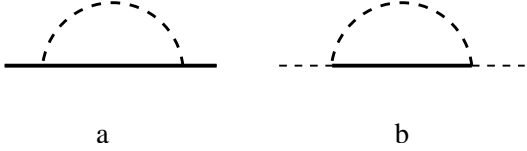,width=7.cm}}
\caption{Self-energy graphs. Solid and dashed lines denote vector mesons (heavy particles)
and Goldstone bosons, respectively. While a) can be treated by the IR method of 
Ref.\protect\cite{BL},
to deal with graphs of type  b), the method developed here has to be used.
\label{fig:self}} 
\end{figure}

There are two cases which have to be distinguished: 1) the momentum $q$ belongs to the heavy particle line of the loop, in the sense that this line is connected to external heavy particle lines (this would necessarily be the case if the heavy particle is a baryon, because of baryon number conservation). For the soft processes we consider, this would mean that 
\begin{displaymath}
q^{2}-M_{V}^{2} = O(p) .
\end{displaymath}
The second case is that 2) the loop is connected only to Goldstone boson lines, which is the case for the vector meson contribution to the Goldstone boson self energy (Fig.~\ref{fig:self}b). 
This can not happen if the heavy particle line represents a baryon. Then
\begin{displaymath}
q^{2} = O(p^{2}) .
\end{displaymath}
We first investigate case 1). To this end we use a Feynman representation
\begin{displaymath}
\frac{1}{ab} = \int_{0}^{1} dz \frac{1}{(a(1-z)+bz)^{2}}
\end{displaymath}
to write
\begin{eqnarray}
&&I_{V\phi}= \nonumber \\
&&i\int \frac{d^{d}k}{(2\pi)^{d}}\int_{0}^{1}dz \frac{1}{[((k-q)^{2}-M_{\phi}^{2})(1-z) + (k^{2}-M_{V}^{2})z]^{2}}\nonumber\\ &&
\end{eqnarray}
Note from this expression that, for $z=0$, the integrand is a pure 'soft pole', while for $z=1$ it is a 'hard pole'. We see that the soft pole structure which we want to extract is associated with the region $z\rightarrow 0$.
\noindent
After a shift $k\rightarrow k+q(1-z)$,the denominator of the integrand becomes 
\begin{displaymath}
[k^{2}-A(z)]^{2}
\end{displaymath}
where
\begin{eqnarray}\label{abdef}
A(z) &=& M_{V}^{2}C(z) , \nonumber\\
C(z) &=& bz^{2} - (b+a-1)z + a ,\nonumber\\
a &=& \frac{M_{\phi}^{2}}{M_{V}^{2}}~,~
b  = \frac{q^{2}}{M_{V}^{2}} ,
\end{eqnarray}
and the $d-$dimensional $k-$integral can be done in a standard way to give
\begin{equation}
I_{V\phi}= -\frac{M_{V}^{d-4}}{(4\pi)^{\frac{d}{2}}}\Gamma \left(2-\frac{d}{2}\right)
\int_{0}^{1}dz (C(z))^{\frac{d}{2}-2} .
\end{equation}
This will develop an infrared singularity as $M_{\phi}\rightarrow 0$ for negative enough dimension $d$. From the expression for $C(z)$, we see that this singularity is located at $z=0$, because in that region we will have
\begin{displaymath}
(C(z))^{\frac{d}{2}-2} \rightarrow \biggl(\frac{M_{\phi}}{M_{V}}\biggr)^{d-4}
\end{displaymath}
generating a part non-analytic in the Goldstone boson mass (remember that $d$ can also be fractional). These findings are consistent with the above observation that the region $z\rightarrow 0$ is to be associated with the soft pole structure, coming from the Goldstone boson propagator.
We come to the conclusion that to extract the soft pole contribution, we should extract the part of the loop integral proportional to noninteger powers of the Goldstone boson mass (for noninteger $d$).

Becher and Leutwyler have proposed a way to achieve this \cite{BL}.
They find that the decomposition of the loop integral into a part non-analytic in $a$ (and therefore $M_{\phi}$) and a part regular in $a$ is given by
\begin{displaymath}
I_{V\phi} = I + R ,
\end{displaymath}
where
\begin{eqnarray}\label{IRsplit}
I &=& -\frac{M_{V}^{d-4}}{(4\pi)^{\frac{d}{2}}}\Gamma\left(2-\frac{d}{2}\right)
\int_{0}^{\infty}(C(z))^{\frac{d}{2}-2}dz \nonumber\\
R &=& +\frac{M_{V}^{d-4}}{(4\pi)^{\frac{d}{2}}}\Gamma\left(2-\frac{d}{2}\right)
\int_{1}^{\infty}(C(z))^{\frac{d}{2}-2}dz  .
\end{eqnarray}
From the above remarks it is clear that $R$, which contains the parameter integral starting at $z=1$, will not produce infrared singular terms for any value of the dimension parameter $d$. It will therefore be analytic in the Goldstone boson mass. On the other hand, the so-called 'infrared singular part' $I$ is exactly the part proportional to noninteger powers of the Goldstone boson mass, for noninteger dimension $d$. Moreover, it is shown in \cite{BL} that this part of the loop integral fulfills power counting (as we would have by now expected for the part associated with the soft pole structure, see the discussion of the last section). We will not repeat the proof for these assertions here, because we are going to do a very similar calculation in the next section.

The parameter integrals for $I$ and $R$ do not converge for $d=4$. To give them an unambiguous value, a partial integration is performed to express them through convergent integrals and a part that is divergent for $d=4$, but can be left away by an analytic continuation argument (analytic continuation from negative values of the parameter $d$). For details, see Ref.\cite{BL}. This leads to an explicit representation of $I$ and $R$ for the case of four dimensions.
 
The infrared regularization (IR) scheme can now be implemented by simply dropping the regular part $R$, argueing that it can be taken care of by an appropriate renormalization of the most general effective Lagrangian. This can be done because it is analytic in the small parameter $M_{\phi}$ (and in external momenta). 
The regular part contains the power counting violating terms originating from the 'hard pole' of the loop integral, which have now been abandoned with the dropping of $R$. We are left with the infrared singular part $I$ which obeys power counting. It still contains a pole in $(d-4)$, which can be dealt with using e.g. the (modified) minimal subtraction scheme. Having done this, we have achieved the goal of a finite loop correction where power counting allows to compute correctly the order with which this correction will appear in the perturbation series.  

The method as presented here strongly relies on dimensional regularization. In particular, the separation of the loop integrals into two parts with fractional versus integer powers of $M_{\phi}$ {\em for a fractional dimension parameter} $d$ allows the argument that chiral symmetry (or more specifically, the Ward identities) have to be obeyed by both parts separately. Therefore dropping one part of it (the regular part) is a chirally symmetric procedure because it leaves us with a regularized amplitude that is again chirally symmetric for itself. It is important here that the scheme of dimensional regularization leaves chiral symmetry untouched, because the validity of the Ward identities does not depend on the space--time dimension parameter. Of course one must deal with {\em all} loop integrals occurring in the perturbation series in the same way to keep the physical content of the theory unchanged. The regular part of any loop integral that one computes will have to be dropped.

If one lets $d \rightarrow$ integer $n$, we will get terms proportional to
\begin{displaymath}
\biggl(\frac{M_{\phi}}{M_{V}}\biggr)^{n+\epsilon} = \biggl(\frac{M_{\phi}}{M_{V}}\biggr)^{n}\biggl(1+\epsilon \ln \biggl(\frac{M_{\phi}}{M_{V}}\biggr)+\ldots \biggr).
\end{displaymath}
In the complete expression, one will leave out terms of $O(\epsilon)$, so that
for an integer dimension, the expression for the infrared part $I$ (up to
$O(\epsilon)$) may well contain a piece analytic in the Goldstone boson mass. 
Only after the separation in the two pieces of different analyticity character 
has been done, it is allowed to let $d$ approach an integer value. This is 
why dimensional regularization, permitting noninteger valued dimension 
parameters, is essential for this approach. Note that an elegant extension
of infrared regularization to multi-loop graphs is given in Ref.\cite{PL}.

%%%%%%%%%%%%%%%%%%%%%%%%%%%%%%%%%%%%%%%%%%%%%%%%%%%%%%%%%%%%%%%%%%%%%%%%%%%%%%%%%%%%%%%%%%%%%%%%%%
\section{Another case of IR regularization}
\setcounter{equation}{0} 
\label{sec:IRnew}
The last section, where the infrared regularization scheme has been introduced, dealt with the case that the momentum squared $q^{2}$, coming from outside the loop, was of the same order as $M_{V}^{2}$. This is the case, for example, for a diagram contributing to the self-energy of a nearly on-shell baryon with a Goldstone boson loop. This was the case considered in \cite{BL}. However, we have seen in the last chapter that vector mesons can occur as strongly virtual resonance states in processes where only soft Goldstone bosons or photons appear as external particles. If the same loop we considered in the last section is connected only to Goldstone boson lines, the momentum flowing into the loop will then be small. Since this is a Goldstone boson momentum, we must require that the 'regular part' which we want to drop from the regularized amplitude is analytic also in this parameter.

Becher and Leutwyler introduce in their original paper the variable
\begin{equation}
\Omega = \frac{q^{2}-M_{\phi}^{2}-M_{V}^{2}}{2M_{\phi}M_{V}}
\end{equation}
which is $O(1)$ for the processes they examine, which correspond to the first of the two cases we distinguished in the last section. They consider the chiral expansion for fixed $\Omega$. But if
\begin{displaymath}
|q^{2}| \ll M_{V}^{2}, \qquad q^{2}= O(p^{2}),
\end{displaymath}
the variable $\Omega$ will be $O(p^{-1})$ ! This already signals that this case will probably have to be treated differently.

Let us first evaluate the integral directly, for $d=4-\epsilon$. We find (omitting terms of $O(\epsilon)$)
\begin{eqnarray*}
I_{V\phi} &=& -\frac{M_{V}^{d-4}}{(4\pi)^{\frac{d}{2}}}\Gamma\left(2-{\frac{d}{2}}\right)
\int_{0}^{1} dz [b(z-x_{1})(z-x_{2})]^{\frac{d}{2}-2} \\ 
 &=& 2\lambda + \frac{1}{16\pi^{2}} + \frac{1}{16\pi^{2}}\int_{0}^{1} dz \ln (b(z-x_{1})(z-x_{2})) \\ 
 &=& 2\lambda - \frac{1}{16\pi^{2}} \\
&& \quad+ \frac{1}{16\pi^{2}}\biggl(x_{1}\ln \biggl(\frac{x_{1}}{x_{1}-1}\biggr) + x_{2}\ln \biggl(\frac{x_{2}}{x_{2}-1}\biggr)\biggr).
\end{eqnarray*}
Here we have introduced the zeroes of $C(z)$,
\begin{equation}\label{zeros}
x_{1,2} = \frac{b+a-1}{2b} \pm \sqrt{\frac{(b+a-1)^{2}-4ab}{4b^{2}}}
\end{equation}
and the standard notation
\begin{displaymath}
\lambda = \frac{M_{V}^{d-4}}{16\pi^{2}}\biggl(\frac{1}{d-4}-\frac{1}{2}(\ln(4\pi)-\gamma+1)\biggr) .
\end{displaymath}
We have selected the mass of the heavy particle as a natural choice 
for the renormalization scale $\mu$ here (this particular choice also leads to
suppression of higher order divergences that appear in the loop integrals and
should thus be made).
Furthermore we have used $C(z=1) = b(1-x_{1})(1-x_{2}) = 1\,$.
To examine this further we write the expansions
\begin{eqnarray}
x_{1} &=& -\frac{a}{1-a} - \frac{ab}{(1-a)^{3}} -\ldots~, \\
\frac{1}{x_{2}} &=& -\frac{b}{1-a} - \frac{b^{2}}{(1-a)^{3}} - \ldots~, 
\end{eqnarray}
showing that, for $b\rightarrow 0$, $x_{2}$ behaves like
\begin{displaymath}
x_{2} \rightarrow \frac{b+a-1}{b} .
\end{displaymath}
Remember that in the case we consider now, $a$ and $b$ are both small variables from their definition, of chiral order $O(p^{2})$.
Using these expansions, and the relation
\begin{displaymath}
C(z=0) = a = bx_{1}x_{2} ,
\end{displaymath}
we see that $I_{V\phi}$ contains a term non-analytic in the Goldstone boson mass
\begin{displaymath}
I_{V\phi} = \frac{1}{16\pi^{2}}x_{1}\ln(a)  +\ldots,
\end{displaymath}
and that it is analytic in the second small variable $b$ (for $|b| \ll 1$ ).

We want to check the power counting for this case: The non-analytic terms are proportional to $x_{1}$, whose expansion in $a$ and $b$ starts at order $O(p^{2})$. This is the expected chiral order for the integral: The loop integration in 4 dimensions gives 4 powers of small momentum, whereas the Goldstone boson propagator gives $-2$. The hard pole structure is of order $O(1)$ here, because the vector meson appears just as an internal resonance line. So the power counting for the non-analytic terms is fine, as expected.

We remark that these non-analytic terms are also produced when one proceeds after the prescription of \cite{Tang}: Expanding the hard pole structure and integrating term by term, one gets $x_{1}$ as the coefficient of the $\ln a$-terms, order by order.

Now we want to do infrared regularization. But now we remark an important point: We can {\em not} simply take over the formulas of the last section. We note that the expression for $I$ and $R$ will contain pieces non-analytic in the other small variable $b$, which is not small in the case treated in the last section. But the original integral does not contain such a non-analyticity in $b$. We conclude that the extension of the parameter integrals to $z\gg 1$ is responsible for this non-analyticity: Somewhere in that region, we must catch up a pole for $b\rightarrow 0$. So we have the problem that, loosely speaking, the 'regular part' is not regular. We will have to modify the method of \cite{BL} somehow, and find a way to separate off the terms non-analytic in the small variable $b$.
To do this, we will have to examine the nature of the singularity we encounter here. This will be done in the next section.

%%%%%%%%%%%%%%%%%%%%%%%%%%%%%%%%%%%%%%%%%%%%%%%%%%%%%%%%%%%%%%%%%%%%%%%%%%%%%%%%%%%%%%%%%%%%%%%%%%%%%%%%%
\section{Singularities in parameter space}
\setcounter{equation}{0} 
\label{sec:sing}
We will consider an integral which has exactly the same features as the one we need, but is a bit simpler. Let
\begin{equation}
\tilde I = \int_{0}^{1}dz\ (z+a)^{-\frac{3}{2}}(1+bz)^{-\frac{3}{2}}
\end{equation}
where $a$ and $b$ are again small parameters in the sense that $|a|, |b|  \ll 1$.
This will develop a singularity at $z=0$ if $a \rightarrow 0$. What if we extend the integration to infinity?  To examine this we first change variables,
\begin{displaymath}
z = \frac{1}{u} ,
\end{displaymath}
and compute
\begin{equation}
\tilde I = \int_{1}^{\infty}u\ du\ (1+au)^{-\frac{3}{2}}(u+b)^{-\frac{3}{2}} .
\end{equation}   
This shows the close similarity between the '$a$-singularity' at $z=0$ and the '$b$-singularity' at $u=0$. If we extend the integration to $z\rightarrow \infty$, we will pick up a non-analytic contribution from $u=0$. A 'regular part' defined as
\begin{equation}\label{Rtilde}
\tilde R = -\int_{1}^{\infty}dz\ (z+a)^{-\frac{3}{2}}(1+bz)^{-\frac{3}{2}}
\end{equation}
will be analytic in $a$, but not in $b$. This is the situation encountered in the last section, for a slightly different integral. How do we get rid of this non-analyticity?

Remembering what we have learned so far, we know how we can get rid of certain non-analytic terms: We must 'destroy the poles' by expanding the pole structures and integrating term by term.
Let us do this:
\begin{eqnarray*}
\tilde I &=& \int_{0}^{1} dz (z+a)^{-\frac{3}{2}}\biggl(1-\frac{3}{2}bz + \frac{15}{8}(bz)^{2}\pm \ldots \biggr) \\ &=& \int_{0}^{1} dz (z+a)^{-\frac{3}{2}}\sum_{m=0}^{\infty} \frac{\Gamma(-\frac{1}{2})}{\Gamma(-\frac{1}{2}-m)} \frac{(bz)^{m}}{m!} \\ &=& \sum_{m=0}^{\infty} \frac{\Gamma(-\frac{1}{2})}{\Gamma(-\frac{1}{2}-m)}\frac{b^m}{m!}\int_{0}^{1}dz (z+a)^{-\frac{3}{2}}z^{m} .
\end{eqnarray*}
Here we were allowed to interchange summation and integration without changing the value of the integral, because in the interval of integration the expansion of the integrand is absolutely convergent when $a,b$ are smaller than 1.

The dependence on $a$ now resides in the simpler integrals of the coefficients in the expansion. The idea is now to find the 'infrared singular part' of each of these coefficients. If the sum of these infrared singular parts converges, it must be the infrared singular part of the full integral $\tilde I$, because the expansion of the pole structure that we have performed did not change the integral.

To find the infrared singular part of
\begin{equation}
I_{m} = \int_{0}^{1}dz (z+a)^{d}z^{m}
\end{equation}
with an arbitrary parameter $d$ (possibly negative) and an integer $m$, we follow the method of \cite{BL} and scale $z=ay$ to find
\begin{equation}
I_{m} = a^{d+m+1}\int_{0}^{\frac{1}{a}}dy\ (1+y)^{d}y^{m} .
\end{equation}
We would now like to take the upper limit of the integration to infinity. Whenever the extended integral converges, it gives
\begin{displaymath}
\int_{0}^{\infty}dy\ (1+y)^{d}y^{m} = \frac{\Gamma(m+1)\Gamma(-d-(m+1))}{\Gamma(-d)} .
\end{displaymath}
There will be a convergence problem if $m$ is large enough. But the divergence comes from large values of $z$ and has nothing to do with the infrared singular part. To see this in detail, let $K\gg 1$ , $m>0$, and do a partial integration:
\begin{eqnarray*}
\int_{0}^{K}dz\ (z+a)^{d}z^{m} &=& K^{m}\frac{(K+a)^{d+1}}{d+1} \nonumber\\
&-& \int_{0}^{K}\frac{m}{d+1}(z+a)^{d+1}z^{m-1}dz~.
\end{eqnarray*}
Doing this $m$ times, and dropping all terms proportional to
$(K+a)^{d+n}$ with $ n \in \mathbf{N}$ ,
because these are clearly expandable around $a=0$ and will therefore not contribute to the piece non-analytic in $a$, we end up with the same result as above, namely, the 'infrared singular part' of $I_{m}$ is {\em always}
\begin{equation}
I_{m, {\rm IR}} = a^{d+m+1}\frac{\Gamma(m+1)\Gamma(-d-(m+1))}{\Gamma(-d)}.
\end{equation}
This form clearly shows the character of the infrared singular part as being proportional to noninteger powers of $a$ for noninteger parameter $d$.

We have now performed the relevant step to find the infrared singular part of each coefficient in the expansion of $\tilde I$. We insert this in the series, setting $d=-\frac{3}{2}$ for definiteness:
\begin{eqnarray*}
\tilde I &=& \sum_{m=0}^{\infty}\frac{\Gamma(-\frac{1}{2})}{\Gamma(-\frac{1}{2}-m)}\frac{b^{m}}{m!}\int_{0}^{1}dz\ (z+a)^{-\frac{3}{2}}z^{m} \\ &\rightarrow & \sum_{m=0}^{\infty}\frac{\Gamma(-\frac{1}{2})}{\Gamma(-\frac{1}{2}-m)}\frac{\Gamma(m+1)\Gamma(\frac{1}{2}-m)}{\Gamma(\frac{3}{2})}\frac{(ab)^{m}a^{-\frac{1}{2}}}{m!} \\ &=& \tilde I_{{\rm IR}} .
\end{eqnarray*}
%Making use of the well known properties 
%\begin{displaymath}
%\Gamma(x+1) = x\Gamma(x) , \qquad \Gamma(m+1)=m! ,
%\end{displaymath}
One can easily sum the series, 
\begin{eqnarray*}
\tilde I_{{\rm IR}} &=& \frac{4}{\sqrt{a}}\sum_{m=0}^{\infty}(ab)^{m}\biggl((m+1)-\frac{1}{2}\biggr) \\ &=& \frac{4}{\sqrt{a}}\biggl(\frac{1}{(1-ab)^{2}}-\frac{1}{2(1-ab)}\biggr) \\ &=& \frac{2}{\sqrt{a}}\frac{1+ab}{(1-ab)^{2}} .
\end{eqnarray*}
The reason why we have selected the value $d=-\frac{3}{2}$ is that it is very easy to compute the integral $\tilde I$ directly. The part non-analytic at $a\rightarrow 0$ can be read off and is the same as the result just given. We have checked this for various other values of the parameter $d$.

In a 'naive' application of the IR method, one would be tempted to simply take
\begin{eqnarray*}
\tilde I_{{\rm BL}} &=&\int_{0}^{\infty} dz\ (z+a)^{-\frac{3}{2}}(1+bz)^{-\frac{3}{2}} \\
&=& \frac{2}{(1-ab)^{2}}\biggl(\frac{1+ab}{\sqrt{a}}-2\sqrt{b}\biggr) ,
\end{eqnarray*}
which contains a part non--analytic in the second small variable $b$. This is the part we have separated off by our procedure.

We note that this '$b$-singular' part can be extracted by proceeding in exact analogy to the steps just performed: Expand the other pole structure, proportional to a power of $(z+a)$, in the integral $\tilde R$, Eq.(\ref{Rtilde}). The expansion is in powers of $a/z$, which is smaller than one in the pertinent interval of integration.

We can now give the result for arbitrary $d$. The general '$a$-singular part' is
\begin{equation}
\tilde I_{{\rm IR}} = \sum_{m=0}^{\infty}\frac{\Gamma(d+1)\Gamma(-d-(m+1))}{\Gamma(-d)\Gamma(d+1-m)}(ab)^{m}a^{d+1} ,
\end{equation}
while the '$b$-singular part' is
\begin{equation}
\tilde I_{{\rm b}} = \sum_{m=0}^{\infty}\frac{\Gamma(d+1)\Gamma(m-2d-1)}{\Gamma(-d)}\frac{(ab)^{m}}{m!}b^{-(d+1)} .
\end{equation}
In the case $d=-\frac{3}{2}$, the last expression yields indeed
\begin{eqnarray*}
\tilde I_{{\rm b}} &=& \sum_{m=0}^{\infty}\frac{\Gamma(-\frac{1}{2})}{\Gamma(\frac{3}{2})}(m+1)(ab)^{m}\sqrt{b}  
= \frac{-4\sqrt{b}}{(1-ab)^{2}}~ ,
\end{eqnarray*}
which is confirmed by the result of the direct calculation. We have
\begin{equation}
\tilde I_{{\rm BL}} = \tilde I_{{\rm IR}} + \tilde I_{{\rm b}}, 
\end{equation}
where the second part is the one we do not want in a genuine regular part (it will appear in $\tilde R$ because the non-analytic behaviour for $b\rightarrow 0$ is not present in the original integral $\tilde I$, with which we started).

What we will need in the next section is the result for $d=-\epsilon$, where, as always, $\epsilon$ is considered as sufficiently small to allow for the neglecting of terms $O(\epsilon^{2})$. In this case,
\begin{equation}
\tilde I_{{\rm IR}} = \sum_{m=0}^{\infty} \frac{\Gamma(1-\epsilon)\Gamma(-(m+1)+\epsilon)}{\Gamma(\epsilon)\Gamma(1-m-\epsilon)}(ab)^{m}a^{1-\epsilon}\, .
\end{equation}
After some $\Gamma$-function algebra, one gets for the sum
\begin{eqnarray*}
\tilde I_{{\rm IR}} &=& a^{1-\epsilon}(-1-\epsilon) -\epsilon \sum_{m=1}^{\infty}\frac{a(ab)^{m}}{m(m+1)} + O(\epsilon^{2}) \\ &=& a(-1-\epsilon + \epsilon \ln(a)) - \epsilon a \sum_{m=1}^{\infty}\biggl(\frac{1}{m}-\frac{1}{m+1}\biggr)(ab)^{m}\\
&& \qquad  \qquad  \qquad \qquad\qquad \qquad +O(\epsilon^{2}).
\end{eqnarray*}
The series can easily be summed
% by use of
%\begin{displaymath}
%\sum_{m=1}^{\infty}\frac{x^{m}}{m} = - \sum_{m=1}^{\infty}(-1)^{m+1}\frac{(-x)^{m}}{m} = -\ln(1-x) ,
%\end{displaymath}
%giving
\begin{equation}
\tilde I_{{\rm IR}} = -a-2\epsilon a + \epsilon a\ln(a)+\epsilon\biggl(\frac{ab-1}{b}\biggr)\ln(1-ab) .
\end{equation}
Please note that the last term is expandable in $b$, and that we have left out terms of $O(\epsilon^{2})$.

With the very same method, we can also compute the '$b$-singular part' for the case $d=-\epsilon$. The result is
\begin{eqnarray}
\tilde I_{{\rm b}} &=& \frac{1}{2}\biggl(a-\frac{1}{b}\biggr)+\frac{1}{2}\epsilon \biggl(a-\frac{1}{b}\biggr)\ln(b) \nonumber \\
&+&\epsilon\biggl(\frac{1-ab}{b}\biggr)(\ln(1-ab)-1).
\label{Itb}
\end{eqnarray}
We have checked that $\tilde I_{{\rm BL}}$, calculated with the method of Becher and Leutwyler, is again the sum of the '$a$-singular part' and the '$b$-singular part', like it was the case for $d=-\frac{3}{2}$.

We have now achieved a method which allows to split integrals of the form of $\tilde I_{{\rm BL}}$ into an 'infrared singular part', behaving non-analytically as $a\rightarrow 0$, and a part showing such behaviour for $b\rightarrow 0$. Note that this decomposition is unique, and that both parts are of a different analyticity character (for fractional $d$), concerning the small parameters $a$ and $b$, respectively. In the next section, we will see how this method can be applied to the scalar loop integral $I_{V\phi}$.

%%%%%%%%%%%%%%%%%%%%%%%%%%%%%%%%%%%%%%%%%%%%%%%%%%%%%%%%%%%%%%%%%%%%%%%%%%%%%%%%%%%%%%%%%%%%%%%%%%%%%%%
\section{Corrected infrared singular part}
\setcounter{equation}{0} 
\label{sec:IRcorr}
The integral we need for the calculation of $I_{V\phi}$ is
\begin{displaymath}
I = \int_{0}^{1}dz\ (b(z-x_{1})(z-x_{2}))^{\frac{d}{2}-2}.
\end{displaymath}
Extracting a factor
\begin{displaymath}
(-bx_{2})^{\frac{d}{2}-2} = (1-(a+b)+\ldots)^{\frac{d}{2}-2}~,
\end{displaymath}
from the integral, which is expandable in $a$ and $b$, the remainder is of the form of $\tilde I$ treated in the last section, because $x_{1}$ and $x_{2}^{-1}$ are small parameters of $O(p^{2})$ :
\begin{displaymath}
I = (-bx_{2})^{\frac{d}{2}-2}\tilde I'
\end{displaymath}
where
\begin{displaymath}
\tilde I' = \int_{0}^{1}dz\ (z+(-x_{1}))^{\frac{d}{2}-2}(1+(-x_{2})^{-1}z)^{\frac{d}{2}-2}.
\end{displaymath}
Doing the appropriate substitutions in Eq.(\ref{Itb}), the infrared singular part of $I$ becomes
\begin{equation}
I_{{\rm IR}}= x_{1}+\epsilon x_{1}-\frac{\epsilon}{2}x_{1}\ln(a)-\frac{\epsilon}{2}(x_{1}-x_{2})\ln\biggl(1-\frac{x_{1}}{x_{2}}\biggr),
\end{equation}
where we have used $a=bx_{1}x_{2}$.
For completeness, we also rewrite the '$b$-singular part':
\begin{eqnarray}
I_{{\rm b}} &=& \frac{x_{2}-x_{1}}{2}-\frac{\epsilon}{4}(x_{2}-x_{1})\ln(bx_{2}^{2}) 
\nonumber \\
&+& \frac{\epsilon}{2}(x_{2}-x_{1})\biggl(1-\ln\biggl(1-\frac{x_{1}}{x_{2}}\biggr)\biggr).
\end{eqnarray}
As a check, we add it to the infrared singular part:
\begin{equation}
I_{{\rm IR}}+I_{{\rm b}}=z_{0}\biggl(1+\epsilon-\frac{\epsilon}{2}\ln(a)\biggr)-\frac{\epsilon}{4}(x_{1}-x_{2})\ln\biggl(\frac{x_{1}}{x_{2}}\biggr).
\end{equation}
Here we have used the notation
\begin{displaymath}
z_{0} = \frac{x_{1}+x_{2}}{2}
\end{displaymath}
as in Ref.~ \cite{BL}. Again, the sum of the '$a$-singular part' and the '$b$-singular part' is the result for the integral   
\begin{displaymath}
\int_{0}^{\infty}dz (b(z-x_{1})(z-x_{2}))^{-\frac{\epsilon}{2}}
\end{displaymath}
when computing it utilizing standard IR. 
But the correct infrared singular part for our case is only a certain part of it, namely, $I_{{\rm IR}}$.

The part which must be split off here, $I_{{\rm b}}$, vanishes if 
\begin{displaymath}
x_{1}=x_{2} \Rightarrow (b+a-1)^{2}-4ab = 0 ,
\end{displaymath} 
(see Eq.(\ref{zeros})), which is the case for
\begin{displaymath}
q^{2} = (M_{V}\pm M_{\phi})^{2} \equiv q^{2}_{\pm} .
\end{displaymath}
We cannot trust our procedure for $q^{2} > M_{V}^{2}$. Therefore the value $q^{2}_{-}$ should be considered as the point where the standard infrared singular part of \cite{BL} and the representation given here, i.e. $I_{{\rm IR}}$, can be joined together.

We emphasize that the kind of argument we have given here is completely in the spirit of the method of Becher and Leutwyler, in that we examined the analyticity properties of the parameter integrals for a general dimension parameter $d$.

The calculation of the infrared singular part of the scalar loop integral $I_{V\phi}$ can now be completed:
\begin{eqnarray}
I_{V\phi}^{{\rm IR}} &=& -\frac{M_{V}^{d-4}}{(4\pi)^{\frac{d}{2}}}
\Gamma\left(2-\frac{d}{2}\right) \, I_{{\rm IR}}
\nonumber\\ 
&=& 2x_{1}\lambda -\frac{1}{16\pi^{2}}\biggl(x_{1}-x_{1}\ln(a)\nonumber \\
&& -(x_{1}-x_{2})\ln\biggl(1-\frac{x_{1}}{x_{2}}\biggr)\biggr) .
\label{IVphiIR}
\end{eqnarray}
Terms of $O(\epsilon)$ have been omitted. We claim that the difference 
\begin{displaymath}
R' \equiv I_{V\phi}-I_{V\phi}^{{\rm IR}}
\end{displaymath}
is the appropriate regular part. This means that it is expandable in the small parameters $a$ and $b$ around zero. The proof consists of two observations:

1) Both $I_{V\phi}$ and $I_{V\phi}^{{\rm IR}}$ contain the same terms non-analytic in $a$, namely,
\begin{displaymath}
\frac{1}{16\pi^{2}}x_{1}\ln(a) .
\end{displaymath}
The difference is therefore expandable in $a$.

2) $I_{V\phi}$ was expandable in $b$ from the start, whereas $I_{V\phi}^{{\rm IR}}$ is expandable in $b$ {\em by construction}. Therefore the difference is of course also expandable in $b$.

Moreover, $R'$ is unique, because we extracted exactly the part of $I_{V\phi}$ proportional to fractional powers of $a$ for fractional dimension parameter $d$. We conclude that $R'$ is a well-defined regular part, and that it can be absorbed in a renormalization of the LECs of the effective Lagrangian.

We add the remark that the expansion of Eq.(\ref{IVphiIR}) is reproduced by 
using the procedure of \cite{Tang}, i.e. expanding the 'hard pole structure' 
and interchanging summation and integration. We have checked this 
to order $O(p^{8})$, but a formal proof that it will give the same 
result to {\em all} orders is still missing. It seems that both procedures are 
indeed consistent (remember, however, the remarks made at the end of
Sect.~\ref{sec:softhard}). This means that the 'low-energy-portion' of loop integrals is, 
in this sense, unambiguous. 
This result is not really surprising: From the arguments of 
Sect.~\ref{sec:softhard}, it is seen that an integral like 
$I_{{\rm soft}}$ (see eq.(\ref{soft})) is a pure 'soft pole' integral, 
i.e. only involving the pole structure associated with the Goldstone boson 
propagator, and thus having no regular part, while an integral like 
$I_{{\rm hard}}$ is regular in the Goldstone boson mass, and does not contain 
fractional powers of $M_{\phi}$ for any choice of the dimension parameter $d$.

%%%%%%%%%%%%%%%%%%%%%%%%%%%%%%%%%%%%%%%%%%%%%%%%%%%%%%%%%%%%%%%%%%%%%%%%%%%%%%%%%%%%%%%%%%%%%%%%%%
\section{Goldstone boson self-energy}
\setcounter{equation}{0} 
\label{sec:self1}
We are now ready to compute the vector meson contribution of the Goldstone boson self-energy.
We consider first the novel type of diagrams where the heavy mass line appears in the loop, 
see Fig.~\ref{fig:self}b.
%\begin{figure}
%\vspace{4cm}
%\caption{Self-energy contribution. Dashed lines: Goldstone bosons, double line: vector meson}
%\end{figure}
We get for this amplitude
\begin{equation}
\frac{12iG_{V}^{2}}{F^{4}}\delta^{ab} I_{\Sigma} \ ,
\end{equation}
using the notation
\begin{equation}\label{Isigma}
I_{\Sigma}(q) = i\int \frac{d^{d}k}{(2\pi)^{d}}\frac{q^{2}k^{2}-(q\cdot k)^{2}}{(k^{2}-M_{V}^{2}+i\epsilon)((k-q)^{2}-M_{\phi}^{2}+i\epsilon)} .
\end{equation}
This integral may be decomposed in a linear combination of scalar loop integrals by standard techniques:
\begin{equation}\label{Isigmad}
I_{\Sigma}= c_{\phi}I_{\phi}+c_{V}I_{V}+c_{V\phi}I_{V\phi},
\end{equation}
where the coefficients $c_{i}$ are given by
\begin{eqnarray}
c_{\phi} &=& \frac{q^{2}+M_{\phi}^{2}-M_{V}^{2}}{4},\nonumber\\ 
c_{V} &=& \frac{q^{2}-M_{\phi}^{2}+M_{V}^{2}}{4},\nonumber \\ 
c_{V\phi} &=&\frac{4q^{2}M_{V}^{2}-(q^{2}-M_{\phi}^{2}+M_{V}^{2})^{2}}{4}, 
\end{eqnarray}
while the scalar loop integrals are defined as
\begin{eqnarray}\label{Iphi}
I_{\phi}=&i\int \frac{d^{d}k}{(2\pi)^{d}}\frac{1}{k^{2}-M_{\phi}^{2}+i\epsilon} =& 2M_{\phi}^{2}\lambda + \frac{M_{\phi}^{2}}{16\pi^{2}}\ln(a), \\ I_{V} =& i\int \frac{d^{d}k}{(2\pi)^{d}}\frac{1}{k^{2}-M_{V}^{2}+i\epsilon} =& 2M_{V}^{2}\lambda , \label{IV}
\end{eqnarray}
and the scalar loop integral $I_{V\phi}$ is defined in Eq.(\ref{IVphidef}).
We repeat the remark that we use $\mu = M_{V}$ for the renormalization scale. 

What power would we like to have for this self-energy amplitude ? We have one loop integration, two vertices of order $O(p^{2})$ and one Goldstone boson propagator. The vector meson propagator is counted as $O(1)$ here, since the vector meson line is not connected to any external heavy particle lines. So we end up with the 'expected' power
\begin{displaymath}
D(\Sigma) = 4 + 2\times2 - (2 + 0) = 6~,
\end{displaymath}
using Eq.(\ref{chiraldim}).
The word 'expected' was used from a naive point of view, because we are already sophisticated enough to expect that the 'hard pole structure' associated with the vector meson propagator will give the loop integral a high energy contribution that spoils the power counting.

Indeed, using the decomposition in scalar loop integrals and the expansions of the variables $x_{1}$ and $x_{2}$ given in eq.(3.10) and (3.11), respectively, it is straightforward to see that $I_{\Sigma}$ contains the following terms which violate the power counting:
\begin{displaymath}
I_{\Sigma}= \frac{1}{4}M_{V}^{4}\biggl(\lambda(6b-2b^{2}+6ab)+\frac{1}{16\pi^{2}}\biggl(\frac{b}{2}+\frac{ab}{2}-\frac{5}{6}b^{2}\biggr)\biggr) + \ldots,
\end{displaymath}
where the dots stand for terms satisfying the power counting, i.e. they are of order $O(p^{6})$ or higher.

To find the 'infrared singular part' of $I_{\Sigma}$, it is sufficient to find the infrared singular part of each of the scalar loop integrals. This is because the coefficients $c_{i}$ do not contain any fractional powers of $M_{\phi}$ for any dimension parameter $d$.

The infrared singular part of $I_{V\phi}$ has been computed in the last section. The integrand of $I_{V}$ is a pure hard pole structure without any dependence on a small parameter like $a$ or $b$, and will therefore not have an infrared singular part. Finally, the integral $I_{\phi}$ is proportional to a fractional power of $M_{\phi}$, as a direct calculation using dimensional regularization shows, so it has no regular part (it does not contain a hard pole structure which could be expanded).
The 'infrared regularized' self-energy amplitude is thus
\begin{equation}
I_{\Sigma}^{{\rm IR}}= \frac{12iG_{V}^{2}}{F^{4}}\delta^{ab}(c_{\phi}I_{\phi} + c_{V\phi}I_{V\phi}^{{\rm IR}}),    
\end{equation}
where $I_{V\phi}^{{\rm IR}}$ is given in eq.(\ref{IVphiIR}).
Using the expansions of the $x_{i}$ and $\ln(1-y) = -y - {y^{2}}/{2} - \ldots$
($|y| \ <  1$),
it can easily be checked that the 'infrared regularized' amplitude obeys power counting.

Before we go on and apply our modified version of infrared regularization to other graphs, we want to mention one more thing. Using the vector field approach, cf.
App.~\ref{app:vec}, the self-energy graph leads to the expression
\begin{displaymath}
\frac{12iG_{V}^{2}}{M_{V}^{2}F^{4}}\delta^{ab}I'_{\Sigma}\ ,
\end{displaymath}
where now
\begin{displaymath}
I'_{\Sigma} = i\int \frac{d^{d}k}{(2\pi)^{d}}\frac{k^{2}(k^{2}q^{2}-(k\cdot q)^{2})}{(k^{2}-M_{V}^{2}+i\epsilon)((k-q)^{2}-M_{\phi}^{2}+i\epsilon)}.
\end{displaymath}
Subtracting the amplitude computed in the vector field approach from the amplitude computed in the tensor field approach, we get
\begin{eqnarray*}
&&A(W)-A(V) =\frac{12iG_{V}^{2}}{M_{V}^{2}F^{4}}\delta^{ab}i \nonumber\\
&&\times 
\int \frac{d^{d}k}{(2\pi)^{d}}\frac{(M_{V}^{2}-k^{2})(k^{2}q^{2}-(k\cdot q)^{2})}{(k^{2}-M_{V}^{2}+i\epsilon)((k-q)^{2}-M_{\phi}^{2}+i\epsilon)}\\ &&= \frac{12G_{V}^{2}}{M_{V}^{2}F^{4}}\delta^{ab}\int \frac{d^{d}k}{(2\pi)^{d}}\frac{k^{2}q^{2}-(k\cdot q)^{2}}{(k-q)^{2}-M_{\phi}^{2}+i\epsilon}
\end{eqnarray*}
But this is the same result as one would get for a self-energy diagram where the vector meson line is replaced by a contact term interaction
\begin{displaymath}
\frac{G_{V}^{2}}{8M_{V}^{2}}\langle [u_{\mu},u_{\nu}][u^{\mu},u^{\nu}]\rangle,
\end{displaymath}
leading to a four-$\phi$ interaction
\begin{displaymath}
-\frac{G_{V}^{2}}{M_{V}^{2}F^{4}}f^{abk}f^{cdk}\partial_{\mu}\phi^{a}\partial_{\nu}\phi^{b}\partial^{\mu}\phi^{c}\partial^{\nu}\phi^{d} .
\end{displaymath}
This confirms the 'duality' between the vector and the tensor field approach \cite{Eck}.
Note that this result does not depend on the regularization scheme - 
it was derived without even computing the integrals. Since the difference of the 
amplitudes is a pure 'soft pole' Goldstone boson loop diagram, we see that 
the 'hard part' (i.e. the regular part) is representation independent. 
In particular, both descriptions produce the same power counting violating terms.

%%%%%%%%%%%%%%%%%%%%%%%%%%%%%%%%%%%%%%%%%%%%%%%%%%%%%%%%%%%%%%%%%%%%%%%%%%%%%%%%%%%%%%%%%%%%%%%%%%
\section{Triangle graph}
\setcounter{equation}{0} 
\label{sec:tria}

\begin{figure}[htb]
\centerline{\psfig{file=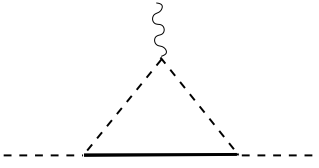,height=2.1cm}}
\caption{Triangle graph as it contributes e.g. to the pion vector form factor.
 Solid, dashed and wiggly lines denote vector mesons (heavy particles),
Goldstone bosons and external sources (fields), respectively. 
\label{fig:tri}} 
\end{figure}

There is one more one-loop diagram of $O(p^{6})$ where the vector meson line 
shows up as a loop line: the triangle diagram of Fig.~\ref{fig:tri}.
%The vertex on top is just the interaction due to the covariant derivatives in $L^{(2)}_{{\rm eff}}$, 
%while the other two vertices follow from the interaction proportional to $G_{V}$. 
\noindent
One gets the following expression for the triangle graph:
\begin{equation}
{\mathcal A}_\Delta
 = \frac{6G_{V}^{2}}{F^{4}}\biggl(f^{ab3}+\frac{1}{\sqrt{3}}f^{ab8}\biggr)I_{\Delta}^{\tau},
\end{equation} 
where the integral is
\begin{eqnarray}
&& -i \, I_{\Delta}^{\tau}(p,p+k)= \int \frac{d^{d}q}{(2\pi)^{d}} \nonumber\\
&&\times\frac{((2p+k) - 2q)^{\tau}
(p\cdot (p+k)q^{2}-(p\cdot q)((p+k)\cdot q))}{(q^{2}-M_{V}^{2})((q-p)^{2}-M_{\phi}^{2})
((q-p-k)^{2}-M_{\phi}^{2})}\, . \nonumber \\
\end{eqnarray}
A decomposition of $I_{\Delta}^{\tau}$ as a linear combination of scalar loop 
integrals is given in App.~\ref{app:int} (for the Goldstone boson momenta on mass shell). 
What concerns us here is the question how the infrared singular part of this integral 
can be obtained. The decomposition in scalar loop integrals contains an 
integral we have not yet treated, namely
\begin{eqnarray}
&&I_{V\phi\phi}(p,p+k)\equiv \int \frac{d^{d}q}{(2\pi)^{d}} \nonumber \\
%&&\equiv \int \frac{d^{d}q}
&&\times \frac{i}{(q^{2}-M_{V}^{2})((q-p)^{2}-M_{\phi}^{2})((q-(p+k))^{2}-M_{\phi}^{2})}~.
\nonumber \\ &&
\end{eqnarray}
Fortunately, it is possible to reduce the problem of finding the infrared singular part of this integral to the case we have already examined. The procedure can in full generality be found in section 6.1 of \cite{BL}. We show how this works in the above example: Introducing one more Feynman parametrization, we write $I_{V\phi\phi}$ as 
\begin{eqnarray*}
&&\int \frac{d^{d}q}{(2\pi)^{d}}\frac{i}{q^{2}-M_{V}^{2}}\int_{0}^{1}dw \times \\&&
\frac{\partial}{\partial M_{\phi}^{2}}\frac{1}{(1-w)((q-p)^{2}-M_{\phi}^{2})+w((q-(p+k))^{2}-M_{\phi}^{2})} \\ &&= \int\frac{d^{d}q}{(2\pi)^{d}}\frac{i}{q^{2}-M_{V}^{2}}\int_{0}^{1}dw \times \\
&&\frac{\partial}{\partial M_{\phi}^{2}}\frac{1}{(q-(p+wk))^{2}-(M_{\phi}^{2}-k^{2}w(1-w))}.
\end{eqnarray*}
The momentum integral is now of the form of $I_{V\phi}$, with the operator
\begin{equation}
\Delta(\ldots) \equiv \int_{0}^{1}dw\ \frac{\partial}{\partial M_{\phi}^{2}}(\ldots)
\end{equation}
acting on it. We can insert our result for $I_{V\phi}^{{\rm IR}}$, with the substitutions
\begin{eqnarray}
a = \frac{M_{\phi}^{2}}{M_{V}^{2}} &\rightarrow& \frac{M_{\phi}^{2}-k^{2}w(1-w)}{M_{V}^{2}} = a'~,
 \\ b=\frac{p^{2}}{M_{V}^{2}} &\rightarrow& \frac{(p+wk)^{2}}{M_{V}^{2}} = b'~.
\end{eqnarray}
Please note that the external Goldstone boson momentum is now called $p$ instead of $q$. 
Also note that the new variables $a'$ and $b'$ are also of $O(p^{2})$, 
which allows to take over the treatment of infrared regularization presented in the foregoing sections.
 
We must show that the operator $\Delta$ does not disturb the properties of infrared singularity 
and power counting. The 'dangerous' part of this operator is the derivative with 
respect to $M_{\phi}^{2}$, since it changes the chiral order. 
It is clear from the above definitions that (for fixed $k^{2}$)
\begin{displaymath}
M_{V}^{2}\frac{\partial}{\partial M_{\phi}^{2}}= \frac{\partial}{\partial a} 
= \frac{\partial}{\partial a'} \ . 
\end{displaymath}
We know from the derivation of the infrared singular part that it can be written in the general form
\begin{displaymath}
I_{V\phi}^{{\rm IR}}(a',b') = (a')^{\frac{d}{2}-1}\sum_{m=0}^{\infty}\sum_{n=0}^{\infty}c_{mn}a'^{m}b'^{n}
\end{displaymath}
with some numerical coefficients $c_{mn}$ that depend only on the dimension $d$. For $d\rightarrow 4$, this gives the correct order $O(p^{2})$ for $I_{V\phi}^{{\rm IR}}$. Letting the operator $\Delta$ act on this expression, we get
\begin{eqnarray}
I_{V\phi\phi}^{{\rm IR}} &=& \Delta I_{V\phi}^{{\rm IR}}(a',b') \nonumber\\ 
&=& \int_{0}^{1}dw\ \frac{1}{M_{V}^{2}}\frac{\partial}{\partial a'}
\biggl((a')^{\frac{d}{2}-1}\sum_{m=0}^{\infty}
\sum_{n=0}^{\infty}c_{mn}a'^{m}b'^{n}\biggr) \nonumber \\ 
&=& \int_{0}^{1}dw\ (a')^{\frac{d}{2}-2}\sum_{m=0}^{\infty}
\sum_{n=0}^{\infty}\left(\frac{d}{2}-1+m\right)c_{mn}a'^{m}b'^{n}~.\nonumber \\&& 
\end{eqnarray}
This shows that the expansion of the thus defined infrared singular part of $I_{V\phi\phi}$ starts with $M_{\phi}^{d-4}$, as one expects for such an integral by simple power counting. We learn from the last expression that, in principle, it is sufficient to know the chiral expansion of $I_{V\phi}^{{\rm IR}}$ to arrive at the chiral expansion of $I_{V\phi\phi}^{{\rm IR}}$. The only problem for practical calculations is that the parameter integrals over $w$ are not at all of a simple form, because $a'$,$b'$ and therefore also $x'_{1}$ and $x'_{2}$, defined in analogy to Eq.(3.9), are nontrivial functions of $w$.

For $d\rightarrow 4$, $I_{V\phi\phi}^{{\rm IR}}$ is
\begin{displaymath}\begin{array}{lll}
\int_{0}^{1}dw\ \frac{1}{M_{V}^{2}}\frac{\partial}{\partial a'}\biggl(2x'_{1}\lambda-\frac{1}{16\pi^{2}}\biggl(x'_{1}-x'_{1}\ln(a')\\-(x'_{1}-x'_{2})\ln(1-\frac{x'_{1}}{x'_{2}})\biggr)\biggr)=\\ \int_{0}^{1} \frac{dw}{M_{V}^{2}}\biggl(2y_{1}\lambda-\frac{1}{16\pi^{2}}\biggl(y_{1}-y_{1}\ln(a') \\- \frac{x'_{1}}{a'}-(y_{1}-y_{2})\ln(1-\frac{x'_{1}}{x'_{2}})-\frac{y_{1}x'_{2}-y_{2}x'_{1}}{x'_{2}}\biggr)\biggr),&
\end{array} \end{displaymath}
where we defined
\begin{displaymath}
y_{1,2}=\frac{\partial}{\partial a'}x'_{1,2} = \frac{1}{2b'}\biggl(1 \pm \frac{b'-a'-1}{\sqrt{(b'+a'-1)^{2}-4a'b'}}\biggr).
\end{displaymath}
The singularity structure of $I_{V\phi\phi}$ is richer than the one of $I_{V\phi}$ because by the definition of $a'$, a term like $\ln(a')$ not only contains the infrared singularity for $M_{\phi}\rightarrow 0$, but also a cut for $k^{2}=t > 4M_{\phi}^{2}$, which is associated with the two-Goldstone boson production threshold.

The important point is that, having the prescription for $I_{V\phi}$, we can find the infrared singular part of any loop integral where a small momentum of order $O(p)$ flows through the heavy particle line(s) (this is the case we have treated in this paper) or where a nearly on-mass-shell heavy particle is involved (in which case the results of \cite{BL} can be used directly). The principle is now well-understood, but practical calculations will be difficult for complicated diagrams, because one needs a parameter integration for every pair of propagators which are combined to one (parameter-dependent) pole structure. An example for this has been shown in the treatment of $I_{V\phi\phi}$. As another point, the decomposition of a complicated loop integral involving a lot of vertices in a decomposition in scalar loop integrals will be lengthy and complicated. But these are no {\em conceptual} problems any more. The conceptual problem of the power-counting violating terms has been solved by dropping the regular parts of all loop integrals, retaining the infrared singular parts that stem from the region where the loop momentum is $O(p)$. As a further consequence, many diagrams, namely those where the loops are formed of heavy particle lines only, can be dropped from the start because the respective loop integrals will only contain 'hard pole' structures and do not lead to an infrared singular part.  
Using this scheme, we can proceed and treat all kinds of diagrams where heavy resonances (not only vector mesons) occur in loops, and whose momenta are to be counted as either nearly on-shell or $O(p)$ by the perturbative scheme of power counting. We finally remark
that the treatment of the vector meson loop graphs in the analysis of the nucleon
electromagnetic form factors performed in \cite{Kubis} is consistent with the procedure
we have established.

%%%%%%%%%%%%%%%%%%%%%%%%%%%%%%%%%%%%%%%%%%%%%%%%%%%%%%%%%%%%%%%%%%%%%%%%%%%%%%%%%%%%%%%%%%%%%%%%%%
\section{Vector meson self-energy graph}
\setcounter{equation}{0} 
\label{sec:Vself}

In the last section we have considered diagrams where the vector meson
 shows up as a strongly virtual intermediate state, with a small momentum 
flowing through the vector meson line. We found the 'infrared singular' part
 of the corresponding amplitude, and we saw that it was necessary to modify
 the method 
of {\cite{BL}}, where all particles in the intermediate state 
where considered as being close to their respective mass shell. 
It is now natural to ask: What happens if the 'light' particles 
(the Goldstone bosons) are far from their mass shell ? In principle, 
the 'hard momentum part' of the pionic intermediate states has been 
integrated out in the effective theory. But in analogy to the case of
 the treatment of the 'heavy' vector meson in the last chapter, it 
might be useful to take these degrees of freedom into account in a
 systematic fashion, because in this way one sums up (infinitely many) 
higher order graphs. We will encounter such a situation in the following section.

\subsection{One more case of IR regularization}
\label{subsec:IRmore}
\begin{figure}[htb]
\centerline{\psfig{file=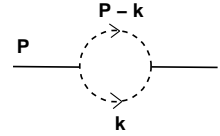,height=1.7cm}}
\caption{The vector meson self-energy diagram with a pure
Goldstone boson loop.  Solid (dashed) lines denote 
vector mesons (Goldstone bosons).
\label{fig:Vself}} 
\end{figure}

In Fig.~\ref{fig:Vself} we show a graph contributing to the vector meson 
self energy. In the case where there is a 'small' ($O(p)$) 
momentum flowing through the vector meson line, the corresponding 
amplitude would be a homogeneous function of small parameters 
(external momentum and quark masses), since the large scale (in this case, 
the vector meson mass) does not show up in the loop line propagators
 and thus cannot produce a 'hard pole' contribution. 
The loop integral is the same as in the Goldstone boson sector, 
and therefore has no 'regular part'.
\noindent
 When computing self-energy contributions, one is usually interested in the
 case where the external momentum $P$ is close to the mass shell 
of the corresponding particle. This leads to the appearance of the large scale 
in the denominator of the integrand of the loop integral, and we expect 
a power-counting violating contribution stemming from a 'hard pole' of the 
integrand. The integral will develop a regular part in the terminology of 
Becher and Leutwyler.

We start our analysis for the case $P^{2} \gg M_{\phi}^{2}$ with the question: 
What is the 'soft part' of a diagram like Fig.4.1? That is, which region of 
the loop momentum integration produces the infrared singular part? Obviously, 
the case where both Goldstone boson propagators are of $O(p^{-2})$ is 
excluded by four-momentum conservation at the two vertices. The region 
where both Goldstone boson lines are far from their mass shell is a pure 
'hard-momentum effect' and thus belongs to the 'regular part'. The 
soft part can only come from the region of the loop integration where 
one line is soft (i.e. it carries an $O(p)$-momentum), and the other 
Goldstone boson line carries the large momentum $P$ and is thus far 
from its mass shell.

As an illustration of this argument, let us first try to extract the 
'soft pole contribution' following the method of {\cite{Tang}}. 
We examine the scalar loop integral
\begin{equation}\label{Idef}
I = i\int \frac{d^{d}k}{(2\pi)^{d}}\frac{1}{(k^{2}-M_{\phi}^{2})((k-P)^{2}-M_{\phi}^{2})}.
\end{equation}
Following the above reflections and the receipt of {\cite{Tang}}, 
we treat the momentum of one line as 'soft' and expand the propagator 
associated with the other line. Then we interchange integration 
and summation of the series, thereby 'destroying the hard pole':
\begin{eqnarray*}
I &\rightarrow& i\int\frac{d^{d}k}{(2\pi)^{d}}\frac{1}{k^{2}-M_{\phi}^{2}}\frac{1}{P^{2}}\sum_{n=0}^{\infty}\frac{(2P\cdot k + M_{\phi}^{2} - k^{2})^{n}}{(P^{2})^{n}}\\ &=& i\int\frac{d^{d}k}{(2\pi)^{d}}\frac{1}{k^{2}-M_{\phi}^{2}}\frac{1}{P^{2}}\sum_{n=0}^{\infty}\frac{(2P\cdot k)^{n}}{(P^{2})^{n}} \\ &\rightarrow& \sum_{n=0}^{\infty} \frac{i}{(P^{2})^{n+1}}\int\frac{d^{d}k}{(2\pi)^{d}}\frac{(2P\cdot k)^{n}}{k^{2}-M_{\phi}^{2}}\\ &=& \frac{1}{2}I_{\rm{soft}}.
\end{eqnarray*} 
The factor of $\frac{1}{2}$ appears due to the fact that the full soft 
part also includes the part where the other line (with momentum $P-k$) 
is considered as 'soft'. Of course this part is equal to the above result 
due to the symmetry of the graph. We note further that it is {\em not} 
legitimate to resum the series in the above result to get
\begin{eqnarray*}
I' &=& i\int
 \frac{d^{d}k}{(2\pi)^{d}}\frac{1}{(k^{2}-M_{\phi}^{2})(P^{2}-2k\cdot P)}
 \\ &=& i\int\frac{d^{d}k}{(2\pi)^{d}}\int_{0}^{\infty}dz 
\\ &\times&\frac{1}{[(k^{2}-M_{\phi}^{2})(1-z)+z((k-P)^{2}-M_{\phi}^{2})]^{2}}
 \\  &=& -i\int\frac{d^{d}k}{(2\pi)^{d}}\int_{1}^{\infty}dz 
\\ &\times&\frac{1}{[(k^{2}-M_{\phi}^{2})(1-z)+z((k-P)^{2}-M_{\phi}^{2})]^{2}}.
\end{eqnarray*}
This contains a 'hard pole' contribution and will not satisfy the power 
counting scheme, which requires the scalar loop integral to be $O(p^{d-2})$, 
because only one Goldstone boson propagator is booked as $O(p^{-2})$, while 
the other Goldstone boson must be far off its mass shell (its momentum must 
be of the order of $P$ by momentum conservation). To repeat, this power
 counting is strictly valid only for the 'soft pole' part of the integral, 
which we have identified as $I_{\rm soft}$. 

The alert reader will note that the result $I_{\rm soft}$ corresponds 
to a series of tadpole graphs, involving only one Goldstone boson propagator. 
This can of course not be the whole story, because the amplitude of 
Fig.~\ref{fig:Vself} has an imaginary part due to the production of two
 Goldstone bosons in the intermediate state, while the tadpole sum does
 not have such an imaginary part. In order to take only $I_{\rm soft}$ as 
the regularized amplitude, one would have to write complex coefficients 
in the effective Lagrangian, which we do not want. A direct calculation 
of the full scalar loop integral shows that the imaginary part does not 
satisfy the power counting mentioned above. This is related to the fact 
that for large $P^{2}$ of the heavy external particle, the Goldstone 
bosons produced in the decay of this particle are not to be considered 
as 'soft'. Below the threshold, we have $P^{2}< 4M_{\phi}^{2}$, so $P^{2}$ 
can not be considered as being very large compared to the 
scale $M_{\phi}^{2}$ in that region, and we would have to take the 
full integral $I$ as the soft part, and not $I_{\rm soft}$.
This phenomenon of the 'missing imaginary part' is consistent with 
the findings of Ref.\cite{bijnens}, where this was noted using the 
Heavy Vector Meson approach. We will not discuss this 
further at this point and turn to the scheme of infrared regularization.
Doing the usual steps, we obtain
\begin{equation}
I = -\frac{\Gamma(2-\frac{d}{2})}{(4\pi)^{\frac{d}{2}}}(P^{2})^{\frac{d}{2}-2}
\int_{0}^{1}dz (D(z))^{\frac{d}{2}-2},
\end{equation}
where
\begin{equation}
D(z) = z^{2}-z+\frac{M_{\phi}^{2}}{P^{2}} .
\end{equation}
Motivated by the remarks made in the last paragraph, we will consider the 
case that $P^{2}> 4M_{\phi}^{2}$. This is fulfilled for the case we are 
interested in, where $P^{2}$ is close to the physical vector meson mass 
squared, and $M_{\phi}$ is the mass of the particles we consider as Goldstone bosons.

Obviously, fractional powers of $M_{\phi}$ are produced in the parameter regions where
\begin{displaymath}
z^{2}-z = 0 \Rightarrow z=0 \ {\rm or}\  z=1,
\end{displaymath}
corresponding to the fact that either one or the other Goldstone boson 
line in the loop carries soft momentum. In accord with the procedure of 
Sect.~\ref{sec:IRnew} (cf. Eq.~(\ref{zeros})), we introduce the zeroes of $D(z)$,
\begin{eqnarray}
d_{1,2} &=& \frac{1}{2}(1 \mp \sigma), \nonumber \\
\sigma &=& \sqrt{1-\frac{4M_{\phi}^{2}}{P^{2}}}~.
\end{eqnarray}
Note that $\sigma \in \mathbf{R}$ and 
\begin{displaymath}
0 < d_{2}-d_{1} = \sigma \leq 1 .
\end{displaymath}
We can simplify our analysis by 'folding' the parameter interval 
symmetrically,\begin{displaymath}
\int_{0}^{1}dz (D(z))^{\frac{d}{2}-2} = 2\int_{0}^{\frac{1}{2}}dz 
(D(z))^{\frac{d}{2}-2},
\end{displaymath}
allowing us to expand the pole due to the zero $d_{2}>\frac{1}{2}$ : 
\begin{eqnarray*}
&&\int_{0}^{1}dz (D(z))^{\frac{d}{2}-2} = \int_{0}^{1}dz
(z-d_{1})^{\frac{d}{2}-2}(z-d_{2})^{\frac{d}{2}-2} \\ &=&
2\sum_{m=0}^{\infty}(-d_{2})^{\frac{d}{2}-2-m}\frac{1}{m!}\frac{\Gamma(\frac{d}{2}-1)}{\Gamma(\frac{d}{2}-1-m)}\\
&&\qquad\times\int_{0}^{\frac{1}{2}}(z-d_{1})^{\frac{d}{2}-2}z^{m} dz .
\end{eqnarray*}
We did not yet change the value of the parameter integral. 
To find the 'infrared singular' part of the parameter integral 
in the last line, we note that $d_{1}$ is proportional to $M_{\phi}^{2}$ 
and of $O(p^{2})$, and shift the integration variable to write
\begin{eqnarray*}
\int_{0}^{\frac{1}{2}}(z-d_{1})^{\frac{d}{2}-2}z^{m}dz &=& 
\int_{-d_{1}}^{0} z^{\frac{d}{2}-2}(z+d_{1})^{m}dz \ \\
&+& \ \int_{0}^{\frac{1}{2}-d_{1}} z^{\frac{d}{2}-2}(z+d_{1})^{m}dz .
\end{eqnarray*}
Terms proportional to $d_{1}^{\frac{d}{2}}$ will only be produced 
by the first term on the right-hand side (remember $m\in \mathbf{N}$).
Scaling the variable of integration with $d_{1}$, it takes the form
\begin{eqnarray*}
&&\int_{-d_{1}}^{0}z^{\frac{d}{2}-2}(z+d_{1})^{m}dz \\
&=&
(-1)^{m+1}(-d_{1})^{\frac{d}{2}-1+m}\int_{0}^{1}t^{\frac{d}{2}-2}(1-t)^{m}dt 
\\ &=& (-1)^{m+1}(-d_{1})^{\frac{d}{2}-1+m}\frac{\Gamma(\frac{d}{2}-1)\Gamma(m+1)}{\Gamma(\frac{d}{2}+m)},
\end{eqnarray*}
where we substituted $z = -td_{1}$.

The 'infrared singular part' of $I$ is thus
\begin{eqnarray}
&&I_{\rm{IR}} =
-\frac{2\Gamma(2-\frac{d}{2})(P^{2})^{\frac{d}{2}-2}}{(4\pi)^{\frac{d}{2}}}
\\&&\times
\sum_{m=0}^{\infty}\frac{(-1)^{m}(d_{1})^{\frac{d}{2}-1+m}(d_{2})^{\frac{d}{2}-2-m}(\Gamma(\frac{d}{2}-1))^{2}}{\Gamma(\frac{d}{2}-1-m)\Gamma(\frac{d}{2}+m)} \, .\nonumber
\end{eqnarray}
This expansion starts with $d_{1}^{\frac{d}{2}-1} \sim M_{\phi}^{d-2}$ and obeys low-energy power counting. The series could be summed up, but this is not necessary. Reviewing what we have done so far, it becomes clear that we have just selected a certain range of integration which produces the fractional powers of $M_{\phi}$. This step may be symbolized by
\begin{eqnarray}
I &=& \int_{0}^{1}dz (\ldots) \rightarrow \int_{0}^{d_{1}}dz (\ldots) 
+ \int_{d_{2}}^{1}dz (\ldots)\nonumber\\
& =& 2\int_{0}^{d_{1}}dz (\ldots) = I_{\rm{IR}}.
\end{eqnarray}
Applying this to $I$ with $d=4-\epsilon$, we find
\begin{equation}\label{VselfIR}
I_{\rm{IR}}= 4d_{1}\lambda+\frac{1}{16\pi^{2}}\biggl(-2d_{1}+\ln(a)+\sigma\ln\biggl(\frac{1+\sigma}{1-\sigma}\biggr)-2\sigma\ln(\sigma)\biggr),
\end{equation}
while the 'regular part' is 
\begin{equation}\label{VselfR}
I-I_{\rm{IR}}= (2-4d_{1})\lambda +\frac{1}{16\pi^{2}}\biggl(-(1-2d_{1})+2\sigma\ln(\sigma)-i\pi\sigma\biggr),
\end{equation}
which is indeed expandable in $M_{\phi}^{2}$ for $P^{2}>4M_{\phi}^{2}$. We have again used $M_{V}$ for the renormalization scale, and the variable $a$ defined in 
Eq.~({\ref{abdef}). 

   It may be checked by expanding $d_{1}$ and $\sigma$ in powers of $M_{\phi}^{2}$ that the infrared singular part indeed satisfies the power counting rules, and also that
\begin{displaymath}
I_{\rm{IR}} = I_{\rm{soft}}.
\end{displaymath}
We have already remarked in the last chapter that the low-energy part of a loop integral is unambiguously defined in this sense.

The imaginary part of the scalar loop integral $I$ is
\begin{displaymath}
                  \frac{-i\sigma}{16\pi},
\end{displaymath}
whose chiral expansion starts $O(1)$ and therefore does not obey the power counting rules. But it cannot be subtracted from the full amplitude since it is not real. The corresponding width of the vector meson due to its possible decay into a pair of Goldstone bosons cannot simply be neglected. In principle, one should give the denominator of the vector meson propagator an imaginary part to deal with this fact.

The result of Eq.~(\ref{VselfIR}) is valid above the two-Goldstone-boson threshold. At
$P^{2}=4M_{\phi}^{2}$, the 'regular part', Eq.~(\ref{VselfR}), vanishes, and remains
zero below the threshold, since as we remarked above the integral $I$ then has
no regular part and is completely 'infrared singular'. The two representations
for the infrared singular part, valid for different ranges of the parameter
$P^{2}$, may be `joined together' at the threshold singularity. A similar
thing happened in the last chapter for the two representations of the infrared
singular part of the scalar loop integral $I_{V\phi}$.

\subsection{Application to the self-energy}
The major problem in finding the 'soft part' of the amplitude of 
Fig.~\ref{fig:Vself} 
has been solved in the last paragraph. For the full expression, we need 
to add some vertex structure from the local effective Lagrangian. 
We choose to work with the interaction Lagrangian of Eq.~(\ref{Wint}) and
refrain from constructing interaction terms with a higher number of 
derivatives, though not all momenta in the present problem can be 
considered as 'soft'. Since the coupling constant $G_{V}$ may be 
measured from $\rho$--meson decay, where the Goldstone bosons are also
not of soft momentum, this can be seen as a valid approximation.
Applying the usual Feynman rules, we obtain
\begin{eqnarray}
&&(-i)\Sigma_{V}^{\mu\nu,\rho\sigma} = \frac{1}{2}\frac{G_{V}^{2}}{F^{4}}
f^{abc}f^{bad} \qquad\nonumber \\
&&\times \int\frac{d^{d}k}{(2\pi)^{d}}\frac{(k^{\mu}P^{\nu}
-k^{\nu}P^{\mu})(P^{\rho}k^{\sigma}-P^{\sigma}k^{\rho})}
{(k^{2}-M_{\phi}^{2})((k-P)^{2}-M_{\phi}^{2})}.
\label{Sigint}
\end{eqnarray}
Before further evaluating this, we have to discuss the 
power counting. The vertices are both of
the order $O(p)$, since only one momentum in the product $k\cdot P$ 
is a small momentum in the sense of the power counting scheme. 
Remembering the discussion of the last paragraph, we want the amplitude 
to be of 'chiral order' $d+1+1-2 = d$. We will see that the infrared 
regularized amplitude will indeed respect this power counting. Using 
the tensor integral of App.~\ref{app:int}, we get
\begin{eqnarray}
&&\Sigma_{V}^{\mu\nu,\rho\sigma} = \\
&&\frac{3G_{V}^{2}}{2F^{4}}
\delta^{cd}P^{\mu\nu,\rho\sigma}\frac{1}{d-1}\biggl(\frac{1}{2}
I_{\phi}+\frac{1}{4}(4M_{\phi}^{2}-P^{2})I\biggr)~. \nonumber
\end{eqnarray}
Here $I$ is the scalar loop integral of Eq.~(\ref{Idef}), and we defined
\begin{equation}\label{P}
P^{\mu\nu,\rho\sigma} = g^{\mu\rho}P^{\nu}P^{\sigma}-g^{\mu\sigma}
P^{\nu}P^{\rho} - (\mu \leftrightarrow \nu)~.
\end{equation}
The infrared regularized amplitude is obtained from (\ref{Sigint}) by simply 
letting $I \rightarrow I_{\rm{IR}}$. The infrared part $I_{\rm{IR}}$ 
was of $O(p^{d-2})$. In order to check that the terms of $O(p^{d-2})$ 
cancel in the soft part of (\ref{Sigint}), it is easiest to use that the first 
term in the chiral expansion of $I_{\rm{IR}}$ is also the first term
 of the series for $I_{\rm{soft}}$, which was given in 
Sect.~\ref{subsec:IRmore}:
\begin{displaymath}
I_{\rm{soft}} = \frac{2}{P^{2}}I_{\phi} + \ldots
\end{displaymath}
Inserting this in $(-i)\Sigma_{V,\rm{IR}}^{\mu\nu,\rho\sigma}$, that is the 
infrared part of Eq.~(\ref{Sigint}), it is clearly seen that the infrared part 
of the amplitude is indeed of order $O(p^{d})$, as required by 
low-energy power counting. 

\subsection{Contributions to the Vector Meson Mass}
First we introduce some notation. We define
\begin{equation}
\mathbf{1} \equiv \mathbf{1}^{\mu\nu,\rho\sigma} = 
\frac{1}{2}(g^{\mu\rho}g^{\nu\sigma}-g^{\mu\sigma}g^{\nu\rho}).
\end{equation}
Furthermore, we write
\begin{displaymath}
\mathbf{P} \equiv P^{\mu\nu,\rho\sigma},
\end{displaymath}
see Eq~(\ref{P}). It is easy to calculate
\begin{eqnarray*}
\mathbf{1}\cdot \mathbf{1} &=& \mathbf{1}~,~~
\mathbf{1}\cdot \mathbf{P} =  \mathbf{P}~, \\
\mathbf{P}\cdot \mathbf{1} &=& \mathbf{P}~,~~
\mathbf{P}\cdot \mathbf{P} = 2P^{2}\mathbf{P}~ ,
\end{eqnarray*}
where the multiplication works as e.g.
\begin{displaymath}
\mathbf{1}_{\mu\nu,\alpha\beta}\cdot\mathbf{P}^{\alpha\beta,\rho\sigma}
 = \mathbf{P}_{\mu\nu}^{\rho\sigma}.
\end{displaymath}
The tensor field propagator may then be written
\begin{equation}
\mathbf{D} = \frac{i}{M_{V}^{2}}\biggl(2\mathbf{1}
+\frac{\mathbf{P}}{M_{V}^{2}-P^{2}}\biggr),
\end{equation}
while its inverse (in the sense of the above multiplication) is
\begin{equation}
\mathbf{D}^{-1} = \frac{1}{i}\biggl(\frac{M_{V}^{2}}{2}\mathbf{1} 
- \frac{1}{4}\mathbf{P}\biggr).
\end{equation}
The one-particle irreducible self-energy amplitude may be parametrized as
\begin{equation}
\mathbf{\Sigma} = \frac{M_{V}^{2}}{2}A\mathbf{1}-\frac{B}{4}\mathbf{P}~.
\end{equation}
where $A$ and $B$ are scalar functions of $P^{2}$ and the meson masses.

The procedure is now standard: Summing over the number of self-energy 
insertions, we find that the full propagator 
\begin{displaymath}
\mathbf{D}_{\rm full} 
= \mathbf{D} + \mathbf{D}(-i)\mathbf{\Sigma}\mathbf{D} + \ldots
\end{displaymath}
is given by
\begin{eqnarray}\label{propfull}
\mathbf{D}_{\rm full} &=& (\mathbf{D}^{-1}+i\mathbf{\Sigma})^{-1}
 \\
&=& \frac{i}{M_{V}^{2}(1-A)}\left(2\mathbf{1}
+\frac{\mathbf{P}}{M_{V}^{2}\biggl(\frac{1-A}{1-B}\biggr)-P^{2}}\right)~.
\nonumber
\end{eqnarray}
We have to look for the poles of this expression. Since $A$ is a 
small perturbation of $O(p^{2})$, the only pole will be at 
\begin{equation}
P^{2} = M_{V}^{2}\biggl(\frac{1-A}{1-B}\biggr)= M_{V,{\rm ph}}^{2}~,
\end{equation}
with $ M_{V,{\rm ph}}$ the physical mass of the vector meson.
Before we use this formula to compute the contribution of 
Fig.~\ref{fig:Vself} to the vector meson mass, 
let us make a very rough estimate of the 
expected size of the contribution. The most general effective 
Lagrangian for the tensor field contains a term
\begin{displaymath}
c\langle W_{\mu\nu}W^{\mu\nu}\chi_{+}\rangle,
\end{displaymath}
yielding, among other terms, a contact term contribution of $O(p^{2})$, 
which gives rise to a shift of the propagator pole:
\begin{displaymath}
M_{V}^{2}\rightarrow M_{V}^{2} + 8cM_{\phi}^{2}~.
\end{displaymath}
Since the coupling constant $c$ is not known, for the purpose of our
 estimate we make a naturalness assumption concerning this coupling, 
and set $c=1$, which gives us a value of 100 MeV for the mass shift. 
If power counting is a consistent perturbative scheme here, we would 
expect for an $O(p^{4})$ correction a number of size of roughly
$(M_{\phi}^{2}/M_{V}^{2})~(100~ \rm{MeV}) \sim 3\,{\rm MeV}$
(for the pion contribution).
Now let us compare this estimate with the 
(infrared regularized) amplitude corresponding to Fig.~\ref{fig:Vself}.
It will contribute to $B$, defined above, with
\begin{eqnarray}
B_{V} &=& -\frac{6G_{V}^{2}}{F^{4}}\biggl(\frac{1}{6}I_{\phi}
+\frac{4M_{\phi}^{2}-M_{V}^{2}}{12}I_{\rm{IR}}\nonumber \\ 
&+& \frac{1}{144\pi^{2}}(d_{1}M_{V}^{2}-(1+4d_{1})M_{\phi}^{2})\biggr),
\end{eqnarray}
giving a mass shift of $1.2\,$MeV, which is really only a small correction,
and also of the size expected by the (very rough) estimate made above.
If we had used the full (real part of) the integral $I$, we would get a
result that is comparable to a correction of $O(p^{2})$ (of course, 
there {\em{are}} such terms of $O(p^{2})$ in the full integral, i.e. 
the power-counting violating terms). We conclude that the main effect 
of the graph of Fig.~\ref{fig:Vself} (at the physical pion mass)
is due to the imaginary part of this diagram, 
associated with the width of the vector meson propagator.

%%%%%%%%%%%%%%%%%%%%%%%%%%%%%%%%%%%%%%%%%%%%%%%%%%%%%%%%%%%%%%%%%%%%%%%%%%%%%%%%%%%%%%%%%%%%%%%%%%
\section{Chiral extrapolation of the rho meson mass}
\setcounter{equation}{0} 
\label{sec:mrho}

In this section, we analyse the quark mass dependence of the $\rho$-meson 
mass and related topics.
This is not entirely new, see e.g. Refs.~\cite{Cohen,Ausrho}, but we do not want
to rely on any model or the assumption of `dominating' contributions to the
$\rho$ self-energy. In fact, there are many different contributions to the
self-energy of the vector mesons, and only a few of the corresponding LECs
are known from phenomenology. Of course, one could resort to models like
the massive Yang-Mills approach or the extended NJL model to estimate these
parameters (as it is done e.g. in the work of Bijnens and collaborators \cite{bijnens}),
but our goal is more modest. We resort to  parameterizing the pion mass dependence of 
$M_\rho$ and fix the combinations of LECs from existing lattice data \cite{CPPACS}.
This allows e.g. to analyze the value of $M_\rho$ in the chiral limit.

First, let us discuss the  many different contributions to the vector meson
mass. We restrict ourselves to terms at most quadratic in the quark masses.
The first type of contribution stems from tree diagrams with quark mass 
insertions, i.e. operators $\sim \chi_+$ or $\sim \chi^2_+$,  like e.g.  
\begin{equation}\label{break}
\langle {\mathbf W} \cdot {\mathbf W} \, \chi_+ \rangle~, 
\langle {\mathbf W} \cdot {\mathbf W} \rangle \langle \chi_+ \rangle~,
\ldots, 
\langle {\mathbf W} \cdot {\mathbf W} \, \chi_+^2 \rangle~, \ldots~.
\end{equation}
The LECs accompanying such explicit symmetry breaking terms are in 
general difficult to determine, as it is well known from the analysis
of the nucleon mass in chiral perturbation theory, see e.g. \cite{FMS,BuM,BL}. 
Such tree
graphs lead to the following vector meson mass terms:\footnote{To avoid
notational clutter, we absorb all prefactors like $1/F^2$ etc. in the
coefficients $k_i$.}
\begin{equation}\label{Mtree}
M_V^{\rm tree} = k_1 \, M_\phi^2  + k_2 \,  M_\phi^4~,
\end{equation}
with $k_1$ ($k_2$) a combination of dimension two (four) LECs.
There is also a tree graph without quark mass insertion, it corresponds
to the vector meson mass in the chiral limited, denoted as $M_V^0$ in what
follows.
Next, we consider the various one-loop graphs.
Tadpole diagrams with an insertions of the second order effective
chiral Lagrangian have also to be considered, some of the pertinent 
structures  are
\begin{eqnarray*}
\langle {\mathbf W} \cdot {\mathbf W} \, \chi_+ \rangle~,  
\langle {\mathbf W}\cdot {\mathbf  W} \, u_\alpha u^\alpha \rangle~, 
\langle  W^{\alpha\mu} W^{\beta\nu} g_{\mu\nu} \, u_\alpha u_\beta \rangle~,
\ldots~.
\end{eqnarray*}
Note that in addition to the symmetry breakers of the type given in
Eq.~(\ref{break}), kinetic terms $\sim \partial_\mu \phi \, \partial^\mu \phi$
from the second order effective Lagrangian also contribute, thus increasing
the number of LECs to be determined. In the comparable case of the nucleon
mass, these can be determined to good accuracy form the analysis of pion-nucleon
scattering in the low energy regime. The total contribution of the tadpoles
to the vector meson mass takes the form
\begin{equation}\label{Mtad}
M_V^{\rm tadpole} =  k_3 \,  M_\phi^4 \, \ln \left( \frac{M_\phi^2}{M_V^2} 
\right)~,
\end{equation}
with $k_3$ another combination of dimension two LECs. The sunrise diagram
(cf. Fig.~\ref{fig:self}a) starts to contribute at order $p^3$ because there
are one derivative vertices of the form 
\begin{displaymath}
\langle \epsilon^{\mu\nu\rho\sigma} \, W_{\mu\nu} \, \nabla^\alpha \,
W_{\alpha\rho} \, u_\sigma \rangle~, \ldots~,
\end{displaymath}
A famous example of such a vertex is the $\omega\rho\pi$ coupling, which is generated
in meson field theory from the Wess-Zumino-Witten term, see e.g. \cite{UlfV,KoichiV}.
It was e.g. considered in the analysis of \cite{Ausrho} as one of what these
authors call `dominating contributions'. Since there are various of such $VV\phi$
couplings, we write the sunrise contribution to the vector meson mass as
\begin{equation}\label{Msun}
M_V^{\rm sunrise} =  k_4 \, M_\phi^3 
+ k_5 \,  M_\phi^4 \, 
\ln \left( \frac{M_\phi^2}{M_V^2} \right) + \ldots ~,
\end{equation}
which is again reminiscent of the leading non-analytic contribution to the
nucleon mass. The ellipsis denotes analytic terms $\sim M_\pi^4$ and higher
order contributions. Finally, we have to consider the self-energy graph considered
in the preceding section. It leads only to a fourth order contribution of
the form
\begin{equation}\label{Mself}
M_V^{\rm self} =  k_6 \,  M_\phi^4 \, \ln \left( \frac{M_\phi^2}{M_V^2} 
\right) + \ldots~,
\end{equation}

To be specific, we consider now the pion mass expansion  of the $\rho$-meson
mass, i.e. we set $M_V = M_\rho$ and $M_\phi = M_\pi$ in the above formulae.
Including {\em only} the non-analytic terms from the fourth order, it takes the form
\begin{equation}\label{extra}
M_\rho = M_\rho^0 + c_1 \, M_\pi^2 + c_2\, M_\pi^3 +
 c_3 \, M_\pi^4 \, \ln \left(\frac{M_\pi^2}{M_\rho^2}\right)
+ {\mathcal O}(M_\pi^4)~,
\end{equation}
where $M_\rho^0$ is the mass in the chiral limit, and the $c_i$ $(i=1,2,3)$ are
combinations of coupling constants as discussed before. In the absence of a
detailed phenomenological analysis of these couplings, we will use the CP-PACS
data \cite{CPPACS} for the $\rho$-meson mass as a function of the pion (average light quark)
mass to determine the parameters $M_\rho^0$, $c_1$, $c_2$ and $c_3$.
We only employ lattice data with $M^2_\pi \lesssim 0.5\,$GeV$^2$. 
In fit~1, we fit these parameters by demanding that the physical
$\rho$-mass is obtained for $M_\pi = 140\,$MeV. For fits~2 and 3, however,
this restriction is lifted. In these
fits, we input the chiral limit mass. Throughout, the fits are subjected
to the further restriction that one obtains
natural values for the combinations of LECs, that is we enforce $|c_i| \leq 3$.
The corresponding fit parameters (obtained by least-square
fits) are collected in Tab.~\ref{tab:1}.

\begin{table}[b]
\caption{Fit parameters. $^\star$ denotes an input quantity. \label{tab:1}}
\begin{center}
\begin{tabular}{|l|ccc|}
\hline
               & ~~Fit~1   & Fit~2   & ~~Fit~3 \\
\hline
$M_\rho^0$ [GeV] & ~~0.776    & ~~0.650$^\star$ & ~~0.800$^\star$ \\
$c_1$ [GeV$^{-1}$] & $-$0.662 & ~~2.200 & $-$1.215 \\
$c_2$ [GeV$^{-2}$] & ~~1.291  & $-$1.934 & ~~1.915 \\
$c_3$ [GeV$^{-3}$] & $-$1.723 & ~~1.572  & $-$2.367 \\
\hline
\end{tabular}
\end{center}
\end{table}
\noindent
\begin{figure}[t]
\centerline{\psfig{file=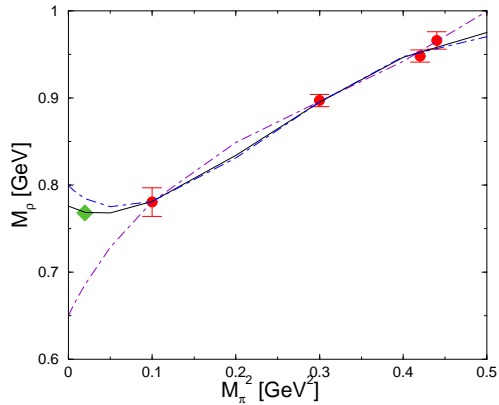,width=6.5cm}}
\caption{The rho meson mass as a function of the light quark mass,
$M_\pi^2 \sim (m_u+m_d)$. The solid (dot-dashed) line(s) refers to fit~1~(2,3)
as described in the text. The lattice data are from CP-PACS \protect
\cite{CPPACS}. The diamond denotes the physical rho mass.
\label{fig:mrho}} 
\end{figure}
\noindent
The corresponding curves are shown in Fig.~\ref{fig:mrho}. To get a
better handle on the theoretical uncertainty, we also allow the fits to
stay within the theoretical uncertainty of the lowest point at $M_\pi^2 =
0.1\,$GeV$^2$, as shown in Fig.~\ref{fig:mrho2}. 
If we insist again on naturalness of the coupling constants,
we can bound the $\rho$-mass in the chiral limit by
\begin{equation}
650~{\rm MeV} \leq M_\rho^0 \leq 800~{\rm MeV}~.
\end{equation}
These results are similar to what was found in the pioneering
work in  Ref.~\cite{Ausrho}, but they are less model-dependent.
The range for $M_\rho^0$ is also consistent with the numbers derived by Bijnens and
collaborators in their study of vector mesons in chiral perturbation theory
\cite{bijnens}. It would be interesting to extend these studies in two
directions, first to include also more recent lattice data and second to
try to give more stringent limits on the combinations of LECs by incorporating
more phenomenological constraints.

\begin{figure}[htb]
\centerline{\psfig{file=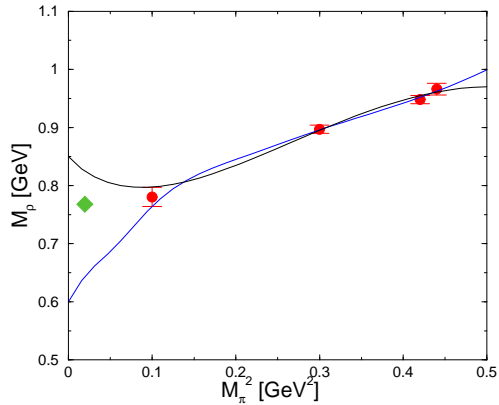,width=6.5cm}}
\caption{The rho meson mass as a function of the pion mass:
Theoretical uncertainty as described in the text. For further
notations, see Fig.~\ref{fig:mrho}.
\label{fig:mrho2}} 
\end{figure}
\noindent

The quark mass expansion of the $\rho$-mass Eq.(\ref{extra}) allows one
to deduce the corresponidng $\sigma$--term,
\begin{equation}
\sigma_{\pi \rho} = \hat m \, \frac{\partial M_\rho}{\partial \hat m} =
M_\pi^2 \,  \frac{\partial M_\rho}{\partial M_\pi^2}~,
\end{equation} 
with $\hat m$ the average light quark mass.
%utilizing $M_\pi^2 = 2B (m_u+m_d)= 2B\hat m$.
From the numbers collected in Table~\ref{tab:1}, we find
\begin{equation}
-1.9\, M_\pi^2 \leq \sigma_{\pi \rho} \leq 1.5\,  M_\pi^2~.
\end{equation} 
This shows again that the rho as a massive particle has a very different
quark mass expansion than the pion, where $\sigma_{\pi} \simeq M_\pi^2$ \cite{GL1}.
In magnitude, the rho $\sigma$-term is similar to the pion-nucleon one, 
$\sigma_{\pi N} \simeq 45\,$MeV.

%%%%%%%%%%%%%%%%%%%%%%%%%%%%%%%%%%%%%%%%%%%%%%%%%%%%%%%%%%%%%%%%%%%%%%%%%%%%%%%%%%%%%%%%%%%%%%%%%%
\section{Summary and outlook}
\setcounter{equation}{0} 
\label{sec:summ}

In this paper, we have considered chiral perturbation theory in the presence
of vector and axial-vector mesons (spin-1) fields and presented an extension
of the infrared regularization scheme originally developed for baryon chiral
perturbation theory. The pertinent results of this investigation can be
summarized as follows:
\begin{itemize}
\item[1)]The most economic way to deal with vector mesons in chiral
  perturbation theory is to utilize the antisymmetric tensor field formulation
  as stressed in \cite{NuB}. When vector mesons appear in tree graphs only,
  calculations are straightforward as summarized in Sect.~\ref{sec:tree} and
  App.~\ref{app:ten}. Of
  course, other formulations like the vector field approach can also be used,
  see  App.~\ref{app:vec},\ref{app:path}.
\item[2)]When vector mesons appear in loops, the appearance of the large mass
  scale complicates the power counting, as discussed in Sect.~\ref{sec:problem}
  and  Sect.~\ref{sec:softhard}. In essence, loop diagrams pick up large
  contributions when the loop momentum is close to the vector meson mass. To the
  contrary, the contribution from the soft poles (momenta of the order of the
  pion mass) that leads to the interesting chiral terms of the low-energy
  EFT (chiral logs and alike) obeys power counting. We have briefly summarized
  the method proposed in \cite{Tang} to extract the `soft pole' contribution from
  one-loop integrals.
\item[3)]The standard case of infrared regularization \cite{BL}, 
  where the heavy particle line is conserved in the (one-loop) graphs
  is recapitulated in Sect.~\ref{sec:IR}. For these cases a very elegant
  splitting of a Feynman parameter integral allows to unambigouosly separate
  the infrared singular from the regular part, cf. Eq.(\ref{IRsplit}).
\item[4)]In the case of spin-1 fields, new classes of self-energy graphs
  appear. The case for lines with small external momenta but a vector meson
  line appearing inside the diagram in analyzed in Sect.~\ref{sec:IRnew} and
  the singularity structure of the corresponding integrals is discussed in
  Sect.~\ref{sec:sing}. In Sect.~\ref{sec:IRcorr} the infrared singular part for such types 
  of integrals is explicitly constructed, cf. Eq.~(\ref{IVphiIR}).
   As explicit examples, the Goldstone
  boson self-energy and the triangle diagram are worked out in Sect.~\ref{sec:self1}
  and Sect.~\ref{sec:tria}, respectively.
\item[5)]A different type of one-loop graphs appears in the vector meson
  self-energy, where only light particles (Goldstone bosons) run in the loop.
  This is discussed in detail in Sect.~\ref{sec:Vself}, where the corresponding
  infrared singular part is extracted, see Eq.~(\ref{VselfIR}), and the  
  contribution to the vector meson mass is worked out. We briefly discuss the
  problems related to the imaginary part of such type of diagrams.
\item[6)]As an application, we consider the pion mass dependence of the $\rho$-meson
  mass in Sect.~\ref{sec:mrho}. We show that there are many contributions with unknown
  LECs, still one is able to derive a compact formula for $M_\rho (M_\pi)$,
  see Eq.~(\ref{extra}). We analyze
  existing lattice data \cite{CPPACS} and conclude that the  $\rho$-meson mass in the
 chiral limit is bounded between 650 and 800~MeV. We have also discussed the
 $\pi\rho$ sigma term.
\end{itemize}

The methods outlined here can be applied to many interesting problems, for example
one could systematically analyze vector meson effects on Goldstone properties like
form factors or polarizabilities or extend these considerations to systems including
baryons (for a first step see e.g. \cite{BM95}).

\section*{Acknowledgements}
We thank Hans Bijnens and J\"urg Gasser for useful comments and
communications.

%%%%%%%%%%%%%%%%%%%%%%%%%%%%%%%%%%%%%%%%%%%%%%%%%%%%%%%%%%%%%%%%%%%%%%
\appendix
\section{Tensor field approach}
\def\theequation{\Alph{section}.\arabic{equation}} 
\setcounter{equation}{0} 
\label{app:ten}

In this appendix, we briefly discuss the representation of spin-1 particles in terms
of antisymmetric tensor fields, following closely Appendix A of \cite{NuB}.

As an antisymmetric tensor field, $W_{\mu \nu}$ has six degrees of freedom, 
whereas a massive vector field only has three. 
Loosely speaking, there are two spin-1 fields 'hidden' in a general 
antisymmetric tensor field approach (corresponding to a reducible representation 
of the rotation group). To make this clear, we decompose the tensor field 
$W_{\mu\nu}$ (in momentum space)
\begin{equation}
W_{\mu\nu}= (W-PWP)_{\mu\nu} + (PWP)_{\mu\nu} \equiv W^{L}_{\mu \nu}+W^{T}_{\mu\nu},
\end{equation}
where the matrix $P$ is the projector
\begin{equation}
P_{\mu\nu}=g_{\mu\nu}-\frac{p_{\mu}p_{\nu}}{p^{2}},
\end{equation}
and $p_{\mu}$ is the momentum four-vector associated with the tensor field.
Because of the projector property of $P$, we have
\begin{displaymath}
p^{\mu}W_{\mu\nu}^{T} = 0,
\end{displaymath}
giving 4 conditions for six degrees of freedom, but one condition is redundant due to the antisymmetry property of $W_{\mu\nu}$. So we are left with $6-3=3$ degrees of freedom for $W^{T}$, and therefore also for $W^{L}$.

Inserting the above decomposition in the general form of an action principle 
for antisymmetric tensor fields \cite{NuB},
\begin{eqnarray}
\tilde S_{W}&=&\int d^{4}x \bigl\{(a-2b)\partial^{\mu}W_{\mu\nu}\partial_{\rho}W^{\rho\nu} + b\partial^{\rho}W_{\mu\nu}\partial_{\rho}W^{\mu\nu} \nonumber \\
&& \qquad\qquad  + cW_{\mu\nu}W^{\mu\nu}\bigr\}~ ,
\end{eqnarray}
with arbitrary parameters $a,b$, and $c\not= 0$, and using
\begin{displaymath}
W_{\mu\nu}^{L}W^{\mu\nu}_{T} = 0,
\end{displaymath}
which is easily verified by a direct calculation, we see that the action splits in two terms:
\begin{equation}
\tilde S_{W} =  S_{W_{L}} + S_{W_{T}},
\end{equation}
where
\begin{eqnarray*}
S_{W_{L}} &=& \int d^{4}x \bigl(a\partial^{\mu}W_{\mu\nu}^{L}\partial_{\rho}W^{\rho\nu}_{L} + cW_{\mu\nu}^{L}W^{\mu\nu}_{L}\bigr), \\ S_{W_{T}} &=& \int d^{4}x \bigl(b\partial^{\rho}W_{\mu\nu}^{T}\partial_{\rho}W^{\mu\nu}_{T} + cW_{\mu\nu}^{T}W^{\mu\nu}_{T}\bigr).
\end{eqnarray*}
Therefore, the path integral can also be factored:
\begin{equation}
\int [dW]e^{i\tilde S_{W}} = \int [dW_{L}]e^{iS_{W_{L}}}\int [dW_{T}]e^{iS_{W_{T}}}.
\end{equation}
In the rest frame, this decomposition corresponds to
\begin{displaymath}
[dW] = [dW_{L}][dW_{T}] = \prod_{i=1}^{3}[dW_{0i}]\prod_{i<j}^{3}[dW_{ij}].
\end{displaymath}
Following \cite{NuB}, we choose $b=0$, so that the second path integral becomes an unimportant constant (from the viewpoint of the classical action, $W_{T}$ becomes a non-propagating field). For $a=0$, in contrast, $W_{L}$ would be 'frozen' in this way.
Choosing, furthermore,
\begin{equation}
a=-\frac{1}{2}, \qquad c=\frac{M_{V}^{2}}{4},
\end{equation}
and dropping the letter $L$, the massive vector field is described by
\begin{eqnarray}
\int[dW]e^{iS_{W}} &=&\int [dW]{\rm exp}\biggl\{i\int d^{4}x\ (-\frac{1}{2}\partial^{\mu}W_{\mu\nu}\partial_{\rho}W^{\rho\nu} \nonumber \\ 
&& \qquad \qquad + \frac{M_{V}^{2}}{4}W_{\mu\nu}W^{\mu\nu})\}.
\end{eqnarray}
Without invalidating the above argument, we can also add a coupling linear in $W$, of the form
\begin{displaymath}
-\frac{1}{4}J_{\mu\nu}W^{\mu\nu},
\end{displaymath}
with some external antisymmetric current $J_{\mu\nu}$.
Since we do not use couplings quadratic in the tensor field in this work, 
we will not attempt to extend the given argument to Lagrangians 
including such more complicated interaction terms.

%%%%%%%%%%%%%%%%%%%%%%%%%%%%%%%%%%%%%%%%%%%%%%%%%%%%%%%%%%%%%%%%%%%%%%
%\appendix
\section{Vector field approach}
\setcounter{equation}{0} 
\label{app:vec}
One can now ask how much the results obtained from calculations similar to 
the ones in the main body of the text depend on the description in terms of an antisymmetric tensor field. Of course a dependence of that kind is not wanted for physical observables!

What happens if one uses a more conventional vector field approach for the description of the vector mesons? 
The Lagrangian for a massive vector field $V_{\mu}$ is well known:
\begin{equation}
{\mathcal L}_{V}^{kin} = -\frac{1}{4}\langle V_{\mu \nu}V^{\mu \nu}\rangle 
+ \frac{1}{2}M_{V}^{2}\langle V_{\mu}V^{\mu}\rangle
\end{equation}
where $V_{\mu \nu}$ is the field strength tensor associated with $V_{\mu}$,
\begin{displaymath}
V_{\mu \nu} = \nabla_{\mu}V_{\nu}-\nabla_{\nu}V_{\mu} .
\end{displaymath}
The corresponding  vector field propagator (in momentum space) reads
\begin{equation}
G_{\mu \nu}(k) = (-i)\frac{\biggl(g_{\mu \nu}-\frac{k_{\mu}k_{\nu}}{M_{V}^{2}}\biggr)}{k^{2}-M_{V}^{2}+i\epsilon}.
\end{equation}
If one now tries to write down interaction terms in analogy to the ones given before, it is seen that they are all $O(p^{3})$, giving a resonance exchange diagram of $O(p^{6})$ in contrast to the $O(p^{4})$ result derived with the antisymmetric tensor field description. In particular, the interaction terms are
\begin{equation}
{\mathcal L}_{V}^{int} = -\frac{f_{V}}{2\sqrt{2}}\langle F^{+}_{\mu \nu}V^{\mu \nu}\rangle 
- \frac{ig_{V}}{2\sqrt{2}}\langle [u_{\mu},u_{\nu}]V^{\mu \nu}\rangle . 
\end{equation}
Here $f_{V}$ and $g_{V}$ are new coupling constants, and the minus sign is a pure convention. This interaction looks very much like Eq.~(\ref{Wint}), but note that the tensor $V_{\mu \nu}$ contains, from its definition, an additional derivative giving the interaction the order $O(p^{3})$ instead of $O(p^{2})$. The form factor contribution derived with this interaction is of higher order than the result of the last section and does {\em not} agree with experiment. If we had started with a vector field description for vector mesons, we might have concluded that the vector meson contribution is less important than suggested, for example, by a dispersive analysis. This problem
was already solved in \cite{Eck}, we simply repeat here some of salient ingredients
in a slightly different way.

The  mathematical relation between the two variants of the theory was worked out
in \cite{Eck} by imposing e.g. the large momentum transfer constraints on the
pion vector form factor. In App.~\ref{app:path}, we argue that the Lagrangians
\begin{equation}
{\mathcal L}_{W}^{kin} - \frac{1}{4}\langle J_{\mu \nu}W^{\mu \nu}\rangle
\end{equation}
and
\begin{equation}
{\mathcal L}_{V}^{kin} + \frac{1}{4M_{V}}\langle J_{\mu \nu}V^{\mu \nu}\rangle 
- \frac{1}{16M_{V}^{2}}\langle J_{\mu \nu}J^{\mu \nu}\rangle
\end{equation}
indeed give equivalent theories on the level of the path integrals, when  $J_{\mu \nu}$ are some traceless hermitian sources. This equivalence is achieved by simply integrating out the field $W_{\mu \nu}$, which leads to the appearance of contact terms quadratic in the source. The source terms we use are 
\begin{displaymath}
J_{\mu \nu} = -\sqrt{2}(F_{V}F^{+}_{\mu \nu} + iG_{V}[u_{\mu},u_{\nu}]) 
\end{displaymath}
(compare Eq.~(\ref{Wint})) and, assuming the validity of the path-integral argument for the equivalence, we draw the conclusion that equivalence leads to
\begin{equation}
f_{V} = \frac{F_{V}}{M_{V}} , \qquad g_{V} = \frac{G_{V}}{M_{V}}~,
\end{equation} 
in perfect agreement with Ref.~\cite{Eck}.
Furthermore, since the $V_{\mu \nu}$- couplings are of $O(p^{3})$, we conclude that vector meson exchange {\em to lowest order} $O(p^{4})$ can be represented by the contact terms
\begin{eqnarray*}
\frac{1}{16M_{V}^{2}}\langle J_{\mu \nu}J^{\mu \nu}\rangle &=& \left<- \frac{G_{V}^{2}}{8M_{V}^{2}}[u_{\mu},u_{\nu}][u^{\mu},u^{\nu}] \right. \nonumber\\ 
&+& \left. \frac{F_{V}^{2}}{8M_{V}^{2}}F^{+}_{\mu \nu}F^{\mu \nu}_{+} + \frac{iF_{V}G_{V}}{4M_{V}^{2}}F^{+}_{\mu \nu}[u^{\mu},u^{\nu}]\right>~ .
\end{eqnarray*}
Using the definition of the objects $u_{\mu}$ and $F^{+}_{\mu \nu}$ and standard trace relations, the contact terms can be written as
\begin{eqnarray*}
-\frac{1}{16M_{V}^{2}}\langle J_{\mu \nu}J^{\mu \nu}\rangle &=& 
L^{V}_{1}P_{1}+L^{V}_{2}P_{2}+L^{V}_{3}P_{3} \nonumber\\
&& +L^{V}_{9}P_{9}+L^{V}_{10}P_{10}+H^{V}_{11}P_{11}~,
\end{eqnarray*}
with $P_{i}$ the structures of the fourth order meson Lagrangian \cite{GL2}
and furthermore
\begin{eqnarray*}
L^{V}_{1}&=&\frac{G_{V}^{2}}{8M_{V}^{2}}~, \qquad L^{V}_{2}=2L^{V}_{1}~, 
\quad L^{V}_{3}=-6L^{V}_{1} \\
L^{V}_{9} &=&\frac{F_{V}G_{V}}{2M_{V}^{2}}~, \quad L^{V}_{10}
=-\frac{F_{V}^{2}}{4M_{V}^{2}}~, \quad 
H^{V}_{11}= -\frac{F_{V}^{2}}{8M_{V}^{2}}~.
\end{eqnarray*}

%%%%%%%%%%%%%%%%%%%%%%%%%%%%%%%%%%%%%%%%%%%%%%%%%%%%%%%%%%%%%%%%%%%%%%
\section{Duality transformation}
\setcounter{equation}{0} 
\label{app:path}

The action for an antisymmetric tensor field $W_{\mu\nu}$ is
\begin{equation}
S_{W} =  \int d^{4}x\ {\mathcal L}_{W}
\end{equation}
Note that $W_{\mu\nu}$ now stands for the field $W_{\mu\nu}^{L}$, having three 
degrees of freedom, as discussed in App.~\ref{app:ten}. 
Now let us modify this action by a term containing a vector field $V_{\mu}$:
\begin{equation}
{\mathcal L}_{V,W}= {\mathcal L}_{W}-\frac{1}{4}J_{\mu\nu}W^{\mu\nu}+\frac{M_{V}^{2}}{2}\biggl(V_{\mu}-\frac{1}{M_{V}}\partial^{\nu}W_{\nu\mu}\biggr)^{2}~.
\end{equation}
We have also added a source term linear in $W_{\mu\nu}$. The equations of motion (e.o.m.) following from this Lagrangian are
\begin{eqnarray}
W_{\mu\nu}+\frac{1}{M_{V}}(\partial_{\mu}V_{\nu}-\partial_{\nu}V_{\mu}) &=& \frac{1}{2M_{V}^{2}}J_{\mu\nu}, \\
V_{\mu}-\frac{1}{M_{V}}(\partial^{\sigma}W_{\sigma\mu}) &=& 0 .
\end{eqnarray}
Substituting $V_{\mu}$ from the second equation into the first one, the latter becomes the 'old' e.o.m. following from $L_{W}-\frac{1}{4}J\cdot W $ alone. So on the classical level, the additional squared term does not change anything.

Considering now path integrals, we deduce by a Gaussian integration over $V$ that
\begin{equation}\label{path}
\int[dW]e^{i\int d^{4}x (L_{W}-\frac{1}{4}J_{\mu\nu}W^{\mu\nu})} \sim \int[dW]\int[dV]e^{i\int d^{4}x L_{V,W}},
\end{equation}
where the '$\sim$' means 'up to a constant factor'. It must be noted here that the vector field $V$ should be treated like $W$, that is using a constraint to eliminate one degree of freedom. On the classical level, the second e.o.m. says that $V$ has the same number of degrees of freedom as $W$, i.e. three d.o.f. Contributions of the path integral over $V$ deviating from this e.o.m are exponentially damped like $e^{-x^{2}}$ because of the form of the squared term in $L_{V,W}$. Using a constraint on $V$, we will get just another constant prefactor as long as this constraint is linear in the field.

If we now recklessly interchange the order of integrations on the
right-hand-side of 
Eq.~(\ref{path}), define 
\begin{displaymath}
V_{\mu\nu}= \partial_{\mu}V_{\nu}-\partial_{\nu}V_{\mu},
\end{displaymath}
and do the $W$-integration, we get
\begin{eqnarray}
&&\int[dW]e^{i\int d^{4}x\ (L_{W}-\frac{1}{4}J_{\mu\nu}W^{\mu\nu})} \nonumber\\
 &\sim& \int[dV][dW]\exp\biggl\{i\int d^{4}x\ \biggl(\frac{1}{4}M_{V}^{2}W_{\mu\nu}W^{\mu\nu}-\frac{1}{4}J_{\mu\nu}W^{\mu\nu}\nonumber\\
&& \qquad\qquad\qquad\qquad\qquad +\frac{M_{V}^{2}}{2}V_{\mu}V^{\mu}+\frac{M_{V}}{2}V_{\mu\nu}W^{\mu\nu}\biggr)\biggr\}\nonumber \\ 
&\sim& \int[dV]\exp\biggl\{i\int d^{4}x\ \biggl(-\frac{1}{4}V_{\mu\nu}V^{\mu\nu}+\frac{M_{V}^{2}}{2}V_{\mu}V^{\mu} \nonumber\\ 
&& \qquad\qquad\qquad\qquad  +\frac{1}{4M_{V}}J_{\mu\nu}V^{\mu\nu} -\frac{1}{16M_{V}^{2}}J_{\mu\nu}J^{\mu\nu}\biggr)\biggr\}~.\nonumber \\ &&
\end{eqnarray}
This corresponds to a conventional vector field Lagrangian, together with contact terms of the form
\begin{displaymath}
-\frac{1}{16M_{V}^{2}}J_{\mu\nu}J^{\mu\nu} .
\end{displaymath}
If the tensor field $W$ is given in the matrix notation of Eq.~(\ref{WT}), which implies that one has to take the (flavor) trace of every term in the Lagrangian $L_{W}$, the source term linear in the field $W_{\mu\nu}^{a}$ can be taken as
\begin{displaymath}
-\frac{1}{4}\langle W_{\mu\nu}J^{\mu\nu}\rangle = -\frac{1}{4\sqrt{2}}W_{\mu\nu}^{a}\langle T^{a}J^{\mu\nu}\rangle .
\end{displaymath}
If $J_{\mu\nu}$ is traceless and hermitian (which is the case for the couplings we consider in Eq.(\ref{Wint})), we produce the contact terms
\begin{displaymath}
\biggl(\frac{1}{16M_{V}^{2}}\biggr)\frac{1}{\sqrt{2}}\langle J_{\mu\nu}T^{a}\rangle \frac{1}{\sqrt{2}}\langle J^{\mu\nu}T^{a}\rangle = \frac{1}{16M_{V}^{2}}\langle J_{\mu\nu}J^{\mu\nu} \rangle ,
\end{displaymath}
which justifies the form of the contact terms given in the preceding appendix.

A path-integral approach to show the duality of the antisymmetric tensor field and the vector field description has also been presented in \cite{Bij}, but the approach used here is shorter. In principle, we have not done much more than a Legendre transformation to change variables from $\partial W$ to $V$. The $W$-field has been integrated out, leaving as a 'souvenir' only the contact terms. As in the foregoing Appendix, we do not attempt to include also more complicated coupling terms in such a transformation, with the excuse that in this work the contact terms are only needed for the linear interaction of Eq.~(\ref{Wint}).

%%%%%%%%%%%%%%%%%%%%%%%%%%%%%%%%%%%%%%%%%%%%%%%%%%%%%%%%%%%%%%%%%%%%%%
\section{Loop integrals}
\setcounter{equation}{0} 
\label{app:int}

In this Appendix we treat the reduction of general loop integrals 
to linear combinations of scalar loop integrals. The scalar loop 
integrals $I_{\phi}$ and $I_{V}$ have been defined in Eqs.(\ref{Iphi}) and (\ref{IV}), respectively,
while $I_{V\phi}$ was defined in Eq.~(\ref{IVphidef}). 
Writing $q^{2}=(q^{2}-M_{V}^{2})+M_{V}^{2}$, one derives
\begin{eqnarray}
&&i\int \frac{d^{d}q}{(2\pi)^{d}}\frac{q^{2}}{(q^{2}-M_{V}^{2})((q-p)^{2}-M_{\phi}^{2})}
\nonumber \\ &&= I_{\phi}+M_{V}^{2}I_{V\phi}(p).
\end{eqnarray}
Moreover, by Lorentz invariance, we must have
\begin{equation}
i\int \frac{d^{d}q}{(2\pi)^{d}}\frac{q^{\mu}}{(q^{2}-M_{V}^{2})((q-p)^{2}-M_{\phi}^{2})} = p^{\mu}S ,
\end{equation}
because there is no other four-vector than $p^{\mu}$ available here. The scalar $S$ can be found by contracting with $p_{\mu}$:
\begin{eqnarray}
S&=&\frac{i}{2p^{2}}\int \frac{d^{d}q}{(2\pi)^{d}}\frac{q^{2}+p^{2}-(q-p)^{2}}{(q^{2}-M_{V}^{2})((q-p)^{2}-M_{\phi}^{2})}\nonumber \\ 
&=& \frac{1}{2p^{2}}(I_{\phi}-I_{V}+(M_{V}^{2}-M_{\phi}^{2}+p^{2})I_{V\phi}).
\end{eqnarray}
Using both results together, we can compute
\begin{eqnarray}
&&i\int \frac{d^{d}q}{(2\pi)^{d}}\frac{(p\cdot q)q^{\mu}}{(q^{2}-M_{V}^{2})((q-p)^{2}-M_{\phi}^{2})} 
\nonumber \\ &&= \frac{p^{\mu}}{4p^{2}}(a_{\phi}I_{\phi}+a_{V}I_{V}+a_{V\phi}I_{V\phi}),
\end{eqnarray}
where the coefficients are given by
\begin{eqnarray}
a_{\phi} &=& M_{V}^{2}-M_{\phi}^{2}+3p^{2}, \nonumber\\
a_{V} &=& M_{\phi}^{2}-M_{V}^{2}-p^{2}~,~~
a_{V\phi} = a_{V}^{2}.
\end{eqnarray}
These results so far are already sufficient to arrive at the decomposition of Eq.~(\ref{Isigmad}) 
for $I_{\Sigma}$. Another useful result is also derived using Lorentz invariance:
\begin{eqnarray}
&&i\int \frac{d^{d}q}{(2\pi)^{d}}\frac{q^{\mu}q^{\nu}}{(q^{2}-M_{\phi}^{2})((q-p)^{2}-M_{\phi}^{2})}
\nonumber \\
&&= g^{\mu\nu}A(p)+\frac{p^{\mu}p^{\nu}}{p^{2}}B(p)~.
\end{eqnarray}
The coefficients can be found by contracting with $g_{\mu\nu}$ and $p_{\mu}$, 
using the above results and $g_{\mu\nu}g^{\mu\nu}=d$ :
\begin{eqnarray}
(d-1)A(p) &=& \frac{1}{2}I_{\phi}+\frac{1}{4}(4M_{\phi}^{2}-p^{2})I_{\phi\phi}(p) \\
(d-1)B(p) &=& \biggl(\frac{d}{2}-1\biggr)I_{\phi} + \biggl(\frac{d}{4}p^{2}-M_{\phi}^{2}\biggr)I_{\phi\phi}(p).\nonumber
\end{eqnarray}
Here the integral $I_{\phi\phi}(p)$ is obtained from $I_{V\phi}(p)$ by substituting 
$M_{\phi}$ for $M_{V}$.

\medskip\noindent
The last result we have to derive is the integral $I_{\Delta}^{\tau}(p,k)$, 
which we only need for on-shell kinematics , $p^{2}=k^{2}=M_{\phi}^{2}$.
Using the algebraic decomposition
\begin{eqnarray*}
&&\frac{4(p\cdot k)q^{2}-4(p\cdot q)(k\cdot q)}{(q^{2}-M_{V}^{2})(q^{2}-2q\cdot p)(q^{2}-2q\cdot k)} 
\\&&= \frac{4(p\cdot k)M_{V}^{2}-M_{V}^{4}}{(q^{2}-M_{V}^{2})(q^{2}-2q\cdot p)(q^{2}-2q\cdot k)} 
\\&&+ \frac{4(p\cdot k)-q^{2}-M_{V}^{2}}{(q^{2}-2q\cdot p)(q^{2}-2q\cdot k)} 
 -\frac{1}{q^{2}-M_{V}^{2}}
\\&&+\frac{1}{q^{2}-2q\cdot p}+\frac{1}{q^{2}-2q\cdot k}
\\&&+\frac{M_{V}^{2}}{(q^{2}-M_{V}^{2})(q^{2}-2q\cdot p)} 
\\&&+ \frac{M_{V}^{2}}{(q^{2}-M_{V}^{2})(q^{2}-2q\cdot k)}~,
\end{eqnarray*}
the results obtained in this appendix enable us to calculate
\begin{eqnarray}
&&I^{\tau}_{\Delta}(p,k)= \nonumber\\
&&i\int\frac{d^{d}q}{(2\pi)^{d}}\frac{(p+k-2q)^{\tau}((p\cdot k)q^{2}-(p\cdot q)(k\cdot q))}
{(q^{2}-M_{V}^{2})((q-p)^{2}-M_{\phi}^{2})((q-k)^{2}-M_{\phi}^{2})}\nonumber\\ 
&&= (p+k)^{\tau}(d_{V\phi\phi}I_{V\phi\phi}(p,k)+d_{V\phi}I_{V\phi}
\nonumber \\ && \quad\quad +d_{\phi\phi}I_{\phi\phi}(k-p)+d_{\phi}I_{\phi}+d_{V}I_{V})~.
\end{eqnarray}
The coefficients are given by
\begin{eqnarray}
4d_{V\phi\phi} &=& \frac{(M_{\phi}^{2}-M_{V}^{2}+(p\cdot k))(4(p\cdot k)M_{V}^{2}-M_{V}^{4})}{M_{\phi}^{2}+(p\cdot k)} ,\nonumber \\ %%&&\\
4d_{V\phi}&=&\frac{2M_{\phi}^{2}M_{V}^{2}-M_{V}^{4}}{M_{\phi}^{2}}-\frac{M_{V}^{4}-4(p\cdot k)M_{V}^{2}}{M_{\phi}^{2}+(p\cdot k)},\nonumber \\
4d_{\phi\phi}&=&\frac{M_{\phi}^{2}+(p\cdot k)}{d-1}-\frac{4(p\cdot k)M_{V}^{2}-M_{V}^{4}}{M_{\phi}^{2}+(p\cdot k)},\nonumber\\
4d_{\phi}&=&\frac{1}{d-1}-\frac{M_{V}^{2}}{M_{\phi}^{2}},~~
4d_{V} = \frac{M_{V}^{2}}{M_{\phi}^{2}}-1.
\end{eqnarray}
We remind the reader that this is valid only for $p^{2}=k^{2}=M_{\phi}^{2}$. For this case, we have
$I_{V\phi}(k)=I_{V\phi}(p)\equiv I_{V\phi}\,$.

%%%%%%%%%%%%%%%% references %%%%%%%%%%%%%%%%%%%%%%%%%%%%%%%%%%%%%%%%%%%%%%%%%%%%%%%%%%%%%%%%% 

\end{document}